\documentclass{emulateapj}
\usepackage{times}
\usepackage{graphicx}
\usepackage{amssymb}
\usepackage{longtable}

\begin{document}

\title{The structures and total (minor + major) merger histories of massive galaxies up to z $\sim$ 3 in the {\it HST} GOODS NICMOS Survey: A possible solution to the size evolution problem}

\author{Asa F. L. Bluck$^{1*}$, Christopher J. Conselice$^{2}$, Fernando~Buitrago$^{2,3}$, Ruth Gr\"utzbauch$^{2}$, Carlos~Hoyos$^{2}$, Alice~Mortlock$^{2}$, Amanda E. Bauer$^{4}$}

\affiliation{ $^{1}$ Gemini Observatory, Northern Operations Center, 670 N. A`ohoku Place, Hilo, Hawaii 96720, USA \\
$^{2}$ Centre for Astronomy and Particle Theory, School of Physics and Astronomy, University of Nottingham, NG7 2RD, UK \\
$^{3}$ Institute for Astronomy, University of Edinburgh, Royal Observatory, Edinburgh, UK \\
$^{4}$ Australian Astronomical Observatory, PO Box 296, Epping, NSW 1710, Australia \\
\\
[ \bf {Submitted to the Astrophysical Journal in original form on Sept 20 2011. \\
Re-submitted after a positive referee report on Nov 19 2011. } ] }

\email{$^{*}$Email: abluck@gemini.edu}

\begin{abstract}

We investigate the total major ($>$ 1:4 by stellar mass) and minor ($>$ 1:100 by stellar mass) merger history of a population of 80 massive ($M_{*} > 10^{11} M_{\odot}$) galaxies at high redshifts (z = 1.7 - 3). We utilise extremely deep and high resolution HST H-band imaging from the GOODS NICMOS Survey (GNS), which corresponds to rest-frame optical wavelengths at the redshifts probed. We find that massive galaxies at high redshifts are often morphologically disturbed, with a $CAS$ deduced merger fraction $f_{m}$ = 0.23 +/- 0.05 at z = 1.7 - 3. We find close accord between close pair methods (within 30 kpc apertures) and $CAS$ methods for deducing major merger fractions at all redshifts. We deduce the total (minor + major) merger history of massive galaxies with $M_{*} > 10^{9} M_{\odot}$ galaxies, and find that this scales roughly linearly with log-stellar-mass and magnitude range. We test our close pair methods by utilizing mock galaxy catalogs from the Millennium Simulation. We compute the total number of mergers to be (4.5 +/- 2.9) / $<\tau_{m}>$ from z = 3 to the present, to a stellar mass sensitivity threshold of $\sim$ 1:100 (where $\tau_{m}$ is the merger timescale in Gyr which varies as a function of mass). This corresponds to an average mass increase of (3.4 +/- 2.2) $\times 10^{11} M_{\odot}$ over the past 11.5 Gyrs due to merging. We show that the size evolution observed for these galaxies may be mostly explained by this merging.

\end{abstract}


\section{Introduction}

The cold dark matter paradigm for structure formation in the Universe leads naturally to a hierarchical picture of galaxy growth, whereby large objects form from the merging together of smaller objects. As such, observing galaxies in a state of merging becomes desirable both as a probe of galaxy evolution and as a critical test of the $\Lambda$CDM cosmological model (see e.g. Bertone $\&$ Conselice 2009). Many studies to date have probed the merger history of massive galaxies, in particular, up to high redshifts (z $<$ 3) see e.g. Patton et al. (2000), Conselice et al. (2003), Conselice et al. (2007), Rawat et al. (2008), Conselice, Yang $\&$ Bluck (2009), Bluck et al. (2009), Lopez-Sanjuan et al. (2010a,b), and Lopez-Sanjuan et al. (2011). 

At low to intermediate redshifts (0 $<$ z $<$ 1.4) close pair methods and morphological approaches find close accord in estimating the major merger history of massive galaxies (Bluck et al. 2009; Conselice, Yang \& Bluck 2009). At the highest redshifts (z $>$ 1.5), studies have concentrated primarily on close pair methods, as opposed to morphological approaches, due to restrictions on the resolution of imaging of very high redshift objects (Bluck et al. 2009). These studies find rough agreement in identifying a positive evolution of the major merger fraction with redshift, with an estimate for the most massive galaxies (with $M_{*} > 10^{11} M_{\odot}$) of evolution such that $f_{m} \propto (1 + z)^{3.0+/-0.4}$, with no sign of this monotonic increase in merger fraction with redshift abating at higher redshifts (see Bluck et al. 2009, and similar results at lower redshifts in e.g. Rawat et al. 2008, Bridge et al. 2010). For lower mass systems there is an observed peak in the merger fraction history at z $\sim$ 1 - 2, as seen in Conselice et al. (2007), but to date no similar peak is observed for the most massive galaxies in the Universe (with $M_{*} > 10^{11} M_{\odot}$), even out to z = 3. This disagrees with predictions from semi-analytical interfaces with the Millennium Simulation, where a turn around in merger fraction is expected for massive galaxies by z $\sim$ 2 (Bertone \& Conselice 2009). Interestingly, however, the merger rates computed in Bluck et al. (2009) agree comfortably with that predicted from the Millennium Simulation (Bertone \& Conselice 2009). At some point the merger fractions must turn over, but this has yet to be observed or constrained for very massive galaxies.

The total number of major mergers that massive galaxies experience since z $\sim$ 3 has been estimated assuming a merger timescale based on N-body simulations (see Lotz et al. 2008a,b and Lotz et al. 2010) as well as via an empirical measure (Conselice 2009), and a parameterisation of the merger history (see Bluck et al. 2009 and Conselice, Yang $\&$ Bluck 2009). Current estimates suggest that the total number of major mergers ($\sim$ 1:4 by stellar mass) experienced since z = 3 by the most massive galaxies in the Universe (with $M_{*} > 10^{11} M_{\odot}$) is $N_{M}$ = 1.7 +/- 0.5 (Bluck et al. 2009). A similar result to this is found recently in Man et al. (2011) where the authors conclude that there are $N_{M}$ = 1.1 +/- 0.5 major mergers over the same epoch. This implies that there is on average a mass increase of less than a factor of two or so due directly to major merging over the past 12 billion years of massive galaxy evolution. Additionally, there is a growing mass of evidence to suggest that over the same time period massive galaxies grow in effective radii by up to a factor of five by building up stellar matter in their outer regions (see e.g. Trujillo et al. 2007, Buitrago et al. 2008, Cimatti et al. 2008, van Dokkum et al. 2009, Carasco et al. 2010). To some extent spectroscopic measures of the velocity dispersions of a very small subset of these high redshift massive galaxies are beginning to confirm that they are extremely dense and compact (e.g. Cenarro \& Trujillo 2009, Cappellari et al. 2009, Onodera et al. 2010).

If this growth in size is a real effect, and remains as dramatic as described currently in the literature, then a crucial question to address is: {\it how do these massive galaxies grow so much in size whilst growing relatively little in mass?} It is now thought that this growth is largely caused by adding stellar material to the outskirts of massive galaxies, such that the luminosity profiles grow in the outer regions of galaxies leaving their cores predominantly unaffected (see Hopkins et al. 2009, Benzanson et al. 2009, Van Dokkum et al. 2010, Buitrago et al. in prep.). Doubtless major merging will play a role in this growth, which is considered in this paper. However, it is also necessary to explore other possibilities, for example, alternative external influences such as minor mergers (also considered in this paper, Khochfar \& Silk 2006, and Naab et al. 2009), cold gas inflow into the galaxies from the intergalactic medium inciting star formation (Dekel et al. 2009, Ceverino et al. 2010, Conselice et al. in preparation), and intrinsic evolution via, for example, AGN `puffing up' scenarios (e.g. Fan et al. 2008 and Bluck et al. 2011). Although these AGN driven models to provoke size evolution have been argued persuasively against recently (Trujillo, Ferreras \& De la Rosa 2011) there is still a case to be made for the indirect effects of AGN redistributing gas around a massive active galaxy.

Low mass galaxies which become integrated within massive galaxies through minor merging can give rise to a variety of triggers - i.e. igniting star formation and AGN activity, and potentially driving size evolution, as well as increasing the stellar mass of massive galaxies directly through adding stars. The merging process also changes the shape and peak of the mass and luminosity function of galaxies with cosmic time (see e.g. Mortlock et al. 2011). Partly these triggers are caused by the fact that lower mass galaxies are often gas rich, and this influx of cool gas into massive galaxies can provide a source of fuel for AGN and material for star formation (e.g. Ricciardelli et al. 2010). This leads to the need to accurately probe the amount and nature of minor, as well as major, mergers a massive galaxy will experience during its evolution, which has until now been extremely difficult to acheive due to the limitations of depth and resolution of imaging galaxies at high redshifts.

In order to investigate the role of major and minor mergers, and structural evolution in massive galaxies, we combine a variety of different techniques and approaches. In this paper we analyse the structural and morphological properties, and minor merger fraction histories, for a sample of 80 massive ($M_{*} > 10^{11} M_{\odot}$) galaxies imaged in the rest frame optical as part of the HST GOODS NICMOS Survey (GNS) at 1.7 $<$ z $<$ 3 (see Conselice et al. 2011a). In particular, we compute the morphological CAS parameters (Conselice et al. 2003) for our sample and use this to obtain an independent major merger fraction for our high redshift sample of massive galaxies to compare with our close pair derived merger fraction in Bluck et al. (2009). We go on to compare the assymetries ($A$, Conselice et al. 2003) of our massive galaxies with the residual flux fraction (RFF, Blakeslee et al. 2006) from their GALFIT fitting (first performed for our GNS galaxies in Buitrago et al. 2008). Furthermore, we exploit the exceptional resolution and depth of the HST GNS by using statistical close pair methods to investigate the minor merger history for these objects, constructing a minor merger history down to a mass sensitivity of $\sim$ 1 : 100 in stellar mass. This paper gives the first reliably measured total merger histories of massive galaxies over the majority of the age of the Universe (over the past 12 billion years), and considers the question of whether or not merging can ultimately explain size evolution.

This paper is structured as follows: \S 2 describes the data used in this study, with \S 3 outlining our results in detail, including morphological and close pair analyses of our sample of high redshift massive galaxies, and a full description of the methods used throughout this paper. \S 4 provides a discussion of our findings and considers the contribution to size evolution made by major and minor merging. \S 5 highlights our conclusions. Throughout the paper we assume a $\Lambda$CDM Cosmology with: H$_{0}$ = 70 km s$^{-1}$ Mpc$^{-1}$, $\Omega_{m}$ = 0.3, $\Omega_{\Lambda}$ = 0.7, and adopt AB magnitude units.

\section{Data and Observations}

The HST GOODS NICMOS Survey (hereafter GNS) images a total of 8298 galaxies in the F160W (H) band, utilising 180 orbits and 60 pointings of the HST NICMOS-3 camera. These pointings are centred around massive galaxies at 1.7 $<$ z $<$ 3.0, which are each observed to 3 orbits depth. Each tile (with area 51.2$''$ $\times$ 51.2$''$, and pixel scale 0.203$''$/pix) is observed in 6 exposures that combine to form images with a final pixel scale of 0.1$''$/pix, and a point spread function (PSF) of $\sim$ 0.3$''$ at full width half maximum (FWHM). The total area of the survey is $\sim$ 43.7 arcmin$^{2}$. Limiting magnitudes reached are H$_{AB}$ = 26.8 (5$\sigma$) (see Conselice et al. 2011a).

The GNS is complete in stellar mass down to $M_{*}$ = 10$^{9.5}$ M$_{\odot}$ at z = 3 to 5 $\sigma$, making it the largest mass complete sample of massive galaxies obtained to date at high redshifts ( z $>$ 1.5). The wealth of observational data available in the GOODS field allowed us to utilise data from the U to the H band in the determination of photometric redshifts. These are computed via a neural network approach, where spectroscopic redshifts in the field are used to train our codes. Multi-colour stellar population fitting techniques were utilised with a Chabrier initial mass function to estimate stellar masses to an accuracy of $\sim$ 0.2 - 0.3 dex. The issue of thermally pulsating AGB stars in model SED fits for massive high redshift galaxies was explored in detail in Conselice et al. (2007) and found not to be a significant problem for our stellar mass determination, perhaps increasing the errors to $\sim$ 0.3 dex, at the upper end of our initial estimation, which is still comfortably within the tolerance required for the analyses presented in this paper. More information on the GNS can be found in Buitrago et al. (2008), Bluck et al. (2009), Gr\"utzbauch et al. (2011a,b), Mortlock et al. (2011), Bluck et al. (2011), and a complete account, incorporating full details on the data acquisition and analysis, and a link to freely access our preliminary data producs including tables and images, in Conselice et al. (2011a). This paper provides full details on the sample selection, observations, source extraction, data catalogs, and determination of photometric redshifts, rest-frame colours and stellar masses.

The initial galaxy selection, of the 80 extrememly massive galaxies at z = 1.7 - 3 studied in detail in this paper, was designed to detect as much as possible all massive galaxies regardless of the age of their stellar populations, dust obscuration, star formation, or nuclear activity. The full details of this selection are provided in \S 2.2 of Conselice et al. (2011a). To summarize, a variety of techniques were used including color-selection methods, e.g. distant red galaxies (DRGs; Franx et al. 2003; Papovich et al. 2006), IRAC extremely red objects (EROs; Yan et al. 2004), and BzK color-selected galaxies (Daddi et al. 2004, 2007). In addition to these color-selected methods we also utilise catalogues of Lyman-break selected BM/BX objects and high-redshift drop-out submilimetre galaxies. The intention being to optimise the total number of massive galaxies in the GNS. We then compute the stellar masses of these galaxies again using the NICMOS H-band + ACS optical bands to confirm the high stellar mass of these systems.

A discussion of more robust error estimates on stellar masses can be found in Bluck et al. (2009), where it is concluded, through Monte-Carlo simulation, that systematic errors arising out of Poisson errors from the steepness of the stellar mass function (Eddington bias) do not result in a significant infiltration of lower mass objects. Moreover, even if some lower mass galaxies do enter our sample they must have very similar merger properties to the genuinely very massive galaxies. Further discussion on the use of the SExtractor package to produce galaxy catalogs is also provided in Bluck et al. (2009) and in Conselice et al. (2011), with special concern for deblending accuracy and removal of stellar objects and spurious detections. In essence point like (non-extended) sources were removed and later objects without galaxy-like colours were removed. This is, of course, vital to achieve well when utilising close pair statistics approaches. For our sample of massive galaxies, each image was looked at carefully by eye and the source extraction parameters were deduced by optimising deblending accuracy.

By way of comparison to lower redshift data, we utilise the POWIR (Palomar Observatory Wide-field InfraRed) survey (PI C. J. Conselice) throughout this paper. Full details of this can be found in Conselice et al. (2007).

\section{Results}
\subsection{CAS Morphologies}

Morphology is one of the oldest methods used for analysing the properties of galaxies (see back to e.g. Hubble 1926). However, it is certainly not without its difficulties and controversies. The need for automated (computational) approaches which are not influenced by subjective observer biases and assumptions has been profoundly lacking throughout much of the history of astronomy, as well as the need for a way to categorise the morphologies of galaxies in a model independent way. One of the recent attempts to address both of these problems is the $CAS$ (concentration, $C$, asymmetry, $A$, clumpiness, $S$) structural parameters designed by Conselice et al. (2003). These offer a physically motivated, non-parametric, and computational scheme to uniquely position galaxies in a 3-dimensional structural space dependent on the concentration, asymmetry and clumpiness of their light profiles. The basic assumption is that galaxies contain information about their natures and histories embedded within their structures, and furthermore that those structures are, or can be, revealed by their observed two-dimensional spacial distribution of optical light. 

Using our data the clumpiness parameter ($S$) is incalculable due to limitations in the resolution of our images (of PSF $\sim$ 0.3$''$). As such, we constrain our analysis here to the $C$ - $A$ plane (or concentration - asymmetry plane), where these parameters require less high spatial resolution and are in general affected much less by redshift effects. $C$ and $A$ are defined in full detail in Conselice, Yang $\&$ Bluck (2009), as well as early definitions in Conselice et al. (2003). For the purposes of this paper it should be sufficient to comment that $A$ is a measure of the total global asymmetry of a galaxy, based on minimising the total residual left from subtracting a 180 degree rotation from the original galaxy image by allowing the centre of rotation to vary within the code (Conselice et al. 2000). A high numerical value of $A$ indicates a large global asymmetry. $C$ is a probe of the concentration of the light distribution and in particlar is constructed from the logarithm of the ratio of the radii at which 80 \% and 20 \% of the light of the entire galaxy (to the Petrosian radius) is contained. Higher values of $C$ indicate more concentrated light profiles.

The structural data for our sample of high redshift massive galaxies is presented in Table 1. We plot relative frequency histograms of the computed $C$ and $A$ parameters for our high redshift GNS massive galaxies (overplotting the histograms for the lower redshift EGS massive galaxies by way of comparison) in Fig. 1. We plot where these massive galaxies lie in $C$ - $A$ space in Fig. 2. We find that there is evolution with cosmic time such that massive galaxies become more concentrated and less asymmetric (to 5$\sigma$ and 4$\sigma$ confidence respectively) at lower redshifts. The evolution in asymmetry will be considered in detail in the subsequent sections (especially in relation to inferred morphologically determined merger fractions), but it is a quantitative reiteration of the qualitative results found in, for example, Forster-Schreiber et al. (2011) and Elmegreen et al. (2011) whereby massive galaxies at higher redshifts are more frequently irregular, and clumpier than their low redshift counterparts. The evolution in concentration is reflective of a tendency for higher redshift massive galaxies to have low Sersic indices and hence exhibit preferentially disk-like structures. This is considered fully in Buitrago et al. in preparation, and is also alluded to in Weinzirl et al. (2011). Presumably, the major merging experienced by these massive galaxies serves to raise their Sersic indices, through transforming rotationally supported spirals into kinematic-pressure supported ellipticals, via an irregular or morphologocally disturbed and kinematically chaotic phase.

Different morphologies of galaxies separate convincingly in the 3-dimensional $C$ - $A$ - $S$ space in the local Universe, with the most significant division for this paper being the region where major merging galaxies tend to reside. It should be noted here that $CAS$ measured merger fractions (such as those presented in the following sections) are only sensitive to major merging of $>$ 1 : 4 by stellar mass (Conselice 2006, Lotz et al. 2008a,b). The applicability of these methods to higher redshifts is considered for the case of identifying major merging systems in the following sub-section. Also in \S 3.1.2 we explore the $CAS$ determined major merger history of massive galaxies in detail.

In the near future these methods can be applied to WFC3 images from the HST with greater reliability and the potential to compare across different wavebands (see e.g. Conselice et al. 2011b). We do not explore the morphologies of the HST ACS images of our galaxies, even though they are of higher resolution than our NICMOS imaging, for two reasons. First, the ACS imaging is in the rest-frame UV for the redshifts of our galaxies and hence not representative of the underlying established stellar populations' light profiles. Therefore, it is not useful for commenting on global asymmetry, as it is dominated by knots of star formation (e.g. Pannella et al. 2009). And secondly, although most of our population of galaxies are visible in the ACS, many of these show up as point sources or very faint extended sources only (see Conselice et al. 2011a). This is most probably due to our sample of massive galaxies having high dust obscuration (see Bauer et al. 2011). These issues emphasise the need for near infrared study of massive galaxies at high redshifts due to the lack of information acquirable about their established (older) stellar populations in the observed optical, hence rest-frame UV at z = 1.5 - 3. For now, our NICMOS imaging is the best available for this task.

\begin{figure*}
\begin{center}
\includegraphics[width=0.48\textwidth,height=0.48\textwidth]{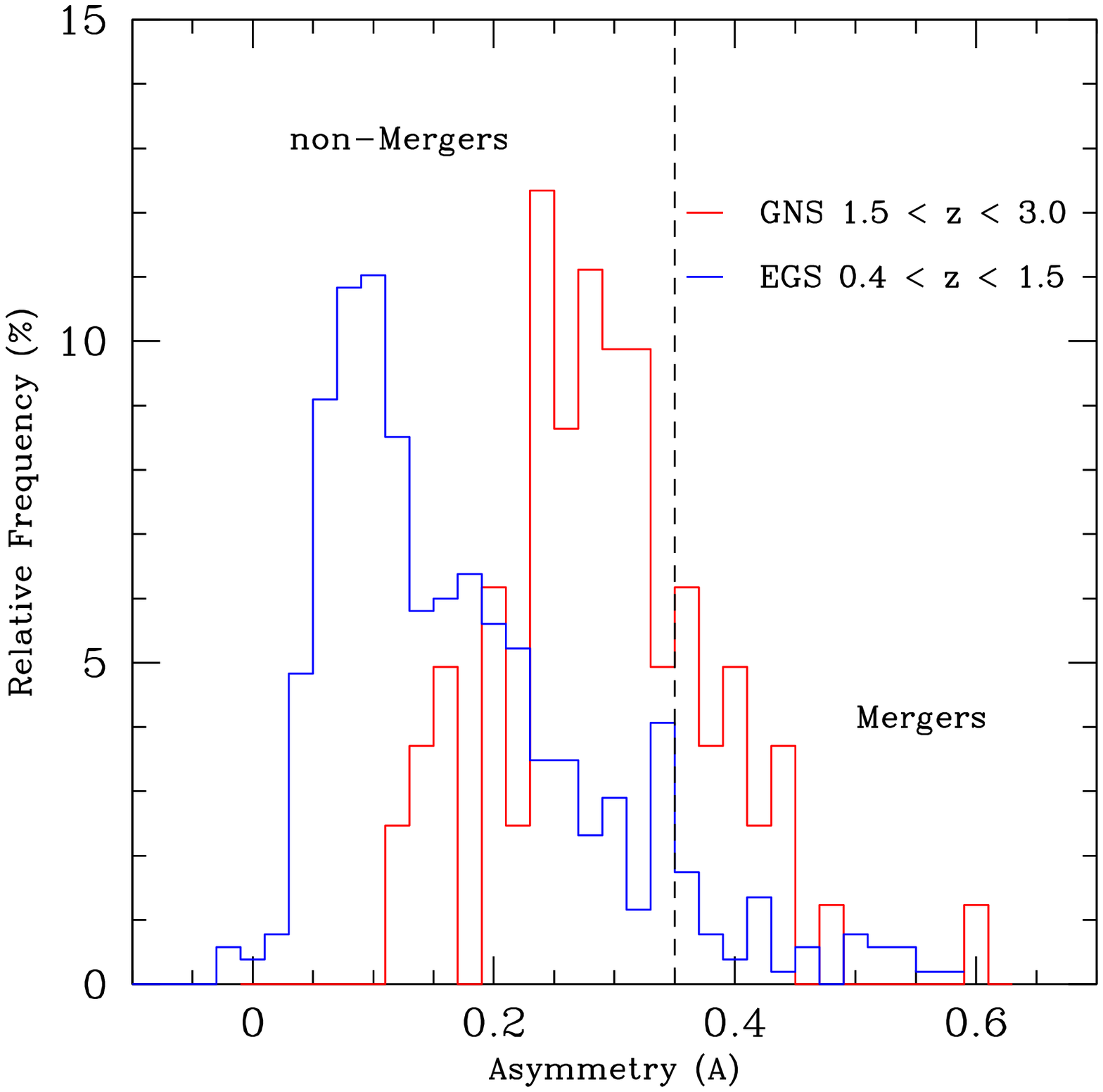}
\includegraphics[width=0.48\textwidth,height=0.48\textwidth]{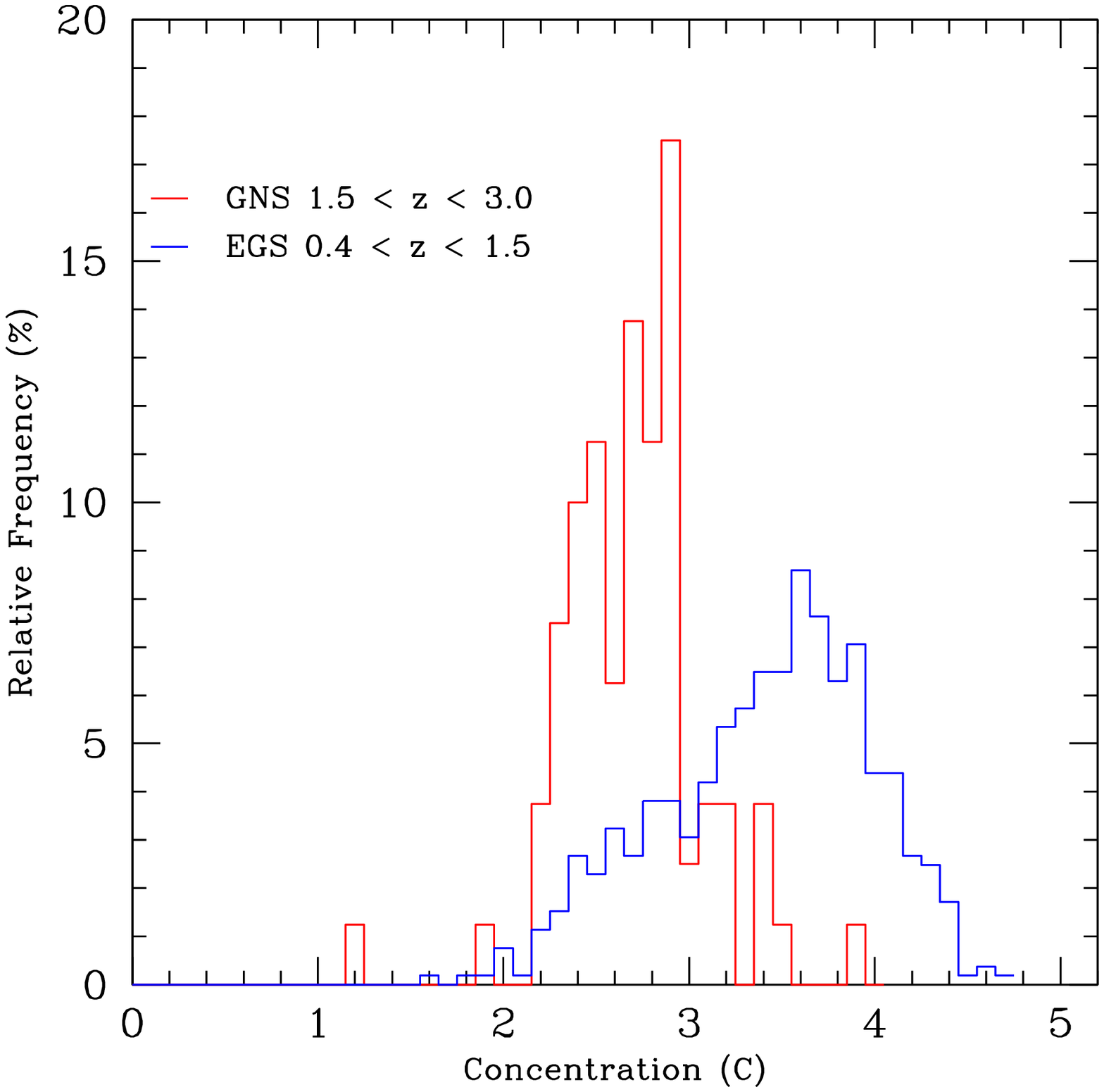}
\caption{Relative frequency histograms for the Asymmetry ($A$, left) and Concentration ($C$, right) parameters. The red (light) histogram is for the high redshift GNS sample of 80 massive ($M_{*} > 10^{11} M_{\odot}$) galaxies at z = 1.7 - 3 taken from the GNS. The blue (dark) histogram is for a comparison set of 524 massive ($M_{*} > 10^{11} M_{\odot}$) galaxies at intermediate redshift (z = 0.4 - 1.5) taken from ACS imaging of POWIR galaxies. There is evolution of the mean asymmetry from $<A>$ = 0.290 +/- 0.010 at z $\sim$ 2.5 to $<A>$ = 0.180 +/- 0.006 at z $\sim$ 1, indicating a 4 $\sigma$ difference from a KS test. There is evolution in the mean concentration from $<C>$ = 2.70 +/- 0.04 at z $\sim$ 2.5 to $<C>$ = 3.40 +/- 0.02 at z $\sim$ 1, indicating a 5 $\sigma$ difference from a further KS test. The errors quoted are the standard error on the mean. Also indicated on the left hand plot are the regions where merging and non-merging galaxies tend to reside for $S$ $<$ $A$ systems.}
\end{center}
\end{figure*}

\begin{figure}
\includegraphics[width=0.47\textwidth,height=0.47\textwidth]{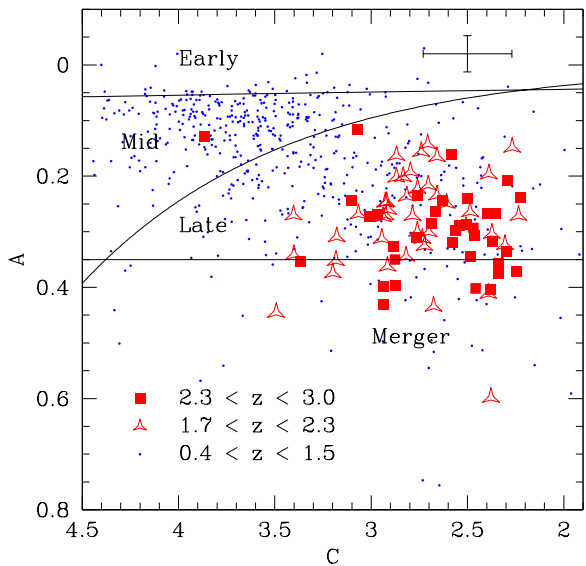} 
\caption[Concentration (C) - Asymmetry (A) Plot]{Asymmetry (A) plotted against Concentration (C) for the 80 massive ($M_{*} > 10^{11} M_{\odot}$) galaxies in our sample at 1.7 $<$ z $<$ 3. The data is divided via redshift, as described on the plot, and the regions where different morphologies reside at z = 0 are also marked on the plot. The placing of almost all non-merging galaxies in the `late type' regime for the two higher redshift ranges is most probably an effect of all of these galaxies having some morphological disturbance (most likely via minor and sub-major merging), and should not be taken too seriously as a morphological classification, since these lines are drawn for z = 0 galaxies. The lowest redshift points (in blue) are taken from the EGS, whereas the higher redshift (red) points are taken from the GNS. The error bar displayed represents the average 1 $\sigma$ errors on the C and A parameters from the CAS code (Conselice et al. 2003).}
\end{figure}

\subsubsection{Using $CAS$ at High Redshifts}

Regions within the $C$-$A$-$S$ space were found to contain almost exclusively certain specific morphological types, e.g. late type galaxies occupy a distinct region of the $C$ - $A$ plane. These distinctions, however, are drawn for local (z $\sim$ 0) galaxies and their extension to the high redshift Universe must be taken with care. Much previous work has aleady looked into this problem, most notably Conselice et al. (2008) and Conselice, Yang \& Bluck (2009). As discussed briefly above, it was concluded that the clumpiness parameter ($S$) is in principle very difficult to compute at high redshifts due to spacial resolution. The concetration ($C$) and asymmetry ($A$) parameters are much more robust at high redshift, but may still contain possible systematic offsets from their true (local) values. Much of the issues associated with using $CAS$ at high redshifts is accounted for if one is careful to study equivalent restframe wavelengths, as done throughout this paper. Nonetheless we run a set of simulations to test possible systematic effects from using $CAS$ at high redshifts.

Our simulations take a broad variety of local Universe galaxy types, from elliptical to irregular galaxies, via a range of spirals. We then simulate how these galaxies would appear in the GNS by degrading the background to the noise level observed in the GNS and dimming the surface brightness in accord with the redshift. The $CAS$ code is then run on both the original image and the artificially simulated high redshift image and any differences are noted. From this analysis we conclude that there is a profound tendency for $S$ to be underestimated at high redsifts, but that the concentration, $C$, and asymmetry, $A$, parameters recover a mean value comfortably within two sigma errors on the $CAS$ code of the original value, where the morphological K-correction is negligible as in the GNS because we view the rest frame optical throughout. 

These simulations are, however, limited in that they assume no surface brightness evolution of galaxies with cosmic time which is known to be false, however, attempts to fully include this are beyond the scope of this first attempt, and moreover should serve to only increase the fidelity of the approach as it would result in brighter (and more easily visible) structures in the light profile at earlier times. There was a slight systematic tendency to underestimate $A$ at high redshifts, but this was only observed at approximately the one sigma level or below. No similar systematic effect with concentration was noted.  

One cause of this possible systematic high redshift effect may be to lower the observed asymmetry ($A$) due to increased background noise, however, our own analysis through simulations of this effect are at best inconclusive. We find that there is perhaps a slight systematic trend to underestimate asymmetry values at high redshifts even after care has been taken to compare equivalent rest frame wavelengths. Therefore, to be secure in our conclusions it is advisable to treat our $A$ values as potential lower estimates at high redshifts, whereas to measure an $A$ value {\it higher} than its true value at high redshifts would be very difficult to achieve. This is intuitive because the low spacial frequency light that $A$ is most sensitive to is smoothed by surface brightness dimming. Future work on interpreting $CAS$ at high redshift is needed, but what follows in this paper is a cautious first attempt, motivated and bolstered by early successes in simulation, and tethered by comparison to other tested methods, i.e. close pairs. Furthermore, the fact that we observe a primary result (that of high redshift massive galaxies being more asymmetric than their local Universe counterparts) in opposition to the dominant systematic effect implies that our conclusions must be robust regardless of this possible systematic.

\subsubsection{Major Merger Fraction from $CAS$ Morphology} 

The classic rest frame optical definition for a $CAS$ selected major ($>$ 1 : 4 by stellar mass) merger is (see Conselice 2003; Conselice, Yang \& Bluck 2009):

\begin{equation}
A > 0.35 \hspace{5mm} \rm{and} \hspace{5mm} A > S
\end{equation}

\noindent That is, a major merging system will have an asymmetry in excess of $A$ = 0.35, and furthermore an asymmetry in excess of its clumpiness, $S$. This selection has a proven track record of cleanly finding major mergers, in the local Universe at least (see Conselice 2003, de Propris et al. 2007). At higher redshifts there are some additional complications (see \S 3.1.1 and Conselice, Yang $\&$ Bluck 2009 for a full discussion). Ultimately, the effect of limitations in the resolution of imaging with the HST NIC-3 camera (PSF $\sim$ 0.3$''$) significantly reduces our sensitivity to the high frequency clumpiness signal in very distant galaxies, in practice setting all $S$ values more or less to zero. This means that the second criteria is irrelevant in practice as if A $>$ 0.35 it is always greater than $S$ at high $z$. This ensures that the asymmetry is a global property of the galaxy and not a result of, for example, clumpy star formation.

Furthermore, because we observe rest frame $V$ to $B$ band with our HST NICMOS H-band images at z = 1.7 - 3, we are probing the visible light and hence stellar population of the galaxies in question, not predominantly star formation as in many other high redshift studies, which frequently view similar high mass high redshift objects in optical light, which corresponds to rest frame UV. For a comparison of some of our galaxies in the NICMOS $H$-band to ACS $z$-band see Fig. 3 in Conselice et al. (2011a). Our massive high-z galaxies are much more clearly visible, with more prominent features, in the H-band, whereas the ACS data is much more difficult to use for morphological classification. The most probable explanation for this is that our sample of massive high-z galaxies are highly dust obscured, see Bauer et al. (2011), as well as having a range of SFRs. The rest frame UV is much more sensitive to knots of patchy star formation and indeed to dust regions causing obscuration than the rest-frame optical. Thus, we have very high resolution imaging (from HST) and additionally see the established stellar population (not star formation) via observing in the near infrared, making the HST GNS an ideal survey with which to probe the structures and morphologies of high redshift massive galaxies. More details on the SEDs of these galaxies is available in Bauer et al. (2011), as well as in Conselice et al. (2011a) and Gruetzbach et al. (2011a,b).

We identify 18 out of 80 massive galaxies that fit the empirical definition of a $CAS$ selected morphologically determined major merger, resulting in a total major merger fraction of $f_{M}$(CAS) = 0.23 +/- 0.05 at 1.7 $<$ z $<$ 3. We further divide this redshift range up into two as in Bluck et al. (2009) to derive a merger fraction of $f_{M}$(CAS) = 0.19 +/- 0.065 (8 out of 43 galaxies) at 1.7 $<$ z $<$ 2.3, and $f_{M}$(CAS) = 0.27 +/- 0.08 (10 out of 37 galaxies) at 2.3 $<$ z $<$ 3. These values are plotted alongside the close pair derived merger fractions for the same galaxies, and the close pair + CAS merger fractions for lower redshift galaxies (from the POWIR Survey, see Conselice et al. 2007) in Fig. 3. Here we add our latest morphologically determined major merger fractions to the merger fraction history plot of Bluck et al. (2009), noting that there is close agreement (to within the 1 $\sigma$ errors) between CAS and close pair (d $<$ 30 kpc) major merger fractions at all redshift ranges probed, including at $z$ $\sim$ 3. This indicates that both approaches are sensitive to the underlying major merger history, with similar detection/ merging timescales (see Conselice, Yang \& Bluck 2009 for a detailed discussion of this issue). However, it should be emphasized that this is an agreement in computing the total number of mergers, not a correspondence of the approaches in indentifying the same galxies. $CAS$ identifies {\it ongoing} major mergers, with the close pair method identifying likely {\it future} mergers. They are in general different populations with some small ($\sim$ 15\%) overlap. See Fig. 4 for typical examples of $CAS$ selected major mergers.

Furthermore, we witness a modest rise in the merger fraction from morphologies, of 8 \% (with $\sim$ 1 $\sigma$ significance) between our two highest redshift ranges, and note that this is only slightly smaller than the observed $\sim$ 2 $\sigma$ increase observed in Bluck et al. (2009) through close pair methods for the same galaxy population. It is possible that at our highest redshift range we are experiencing a systematic effect that effectively smooths our galaxies' light, thus underestimating $A$ slightly. For now it is sufficient to note that broadly speaking there is agreement between close pair and CAS methods of determining major merger fractions, and that in both schemes there is a very high chance (4 $\sigma$ rise in $f_{m}$ from z = 0 to z = 3 in both CAS and close pair methodologies) that the major merger fraction of massive galaxies was considerably higher at high redshifts than in the local Universe (as shown in Fig. 3). At lower redshifts (where there are numerous major merger fraction results for massive galaxies in the literature) our work agrees closely with many other studies using a variety of approaches from close pair statistics to morphologies, including Rawat et al. (2008), Jogee et al. (2009), Conselice, Yang \& Bluck (2009), Bridge et al. (2010), Lopez-Sanjuan et al. (2010a,b). The highest redshift ranges probed here are the first such morphological measurements to the authors' knowledge, but recent complementary results in agreement to these have been presented in Man et al. (2011) using a statistical approach similar to Bluck et al. (2009). 

\begin{center}
\begin{figure}
\includegraphics[width=0.45\textwidth,height=0.45\textwidth]{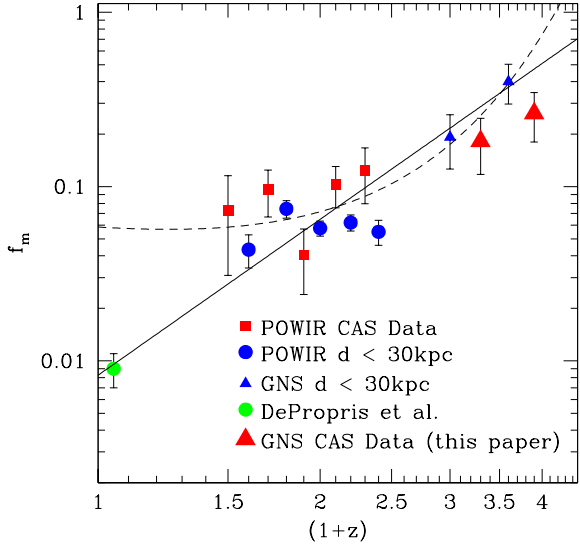} 
\caption{The major merger ($f_{m}$) evolution with redshift for massive ($M_{*} > 10^{11} M_{\odot}$) galaxies. This plot utilises data from Bluck et al. (2009) and Conselice et al. (2007), where the major merger history of massive galaxies is probed using statistical close pair methods (at all z) and CAS morphologies (at z $<$ 1.5). In this paper we add the CAS morphology determined major merger fractions for the two highest redshift ranges, up to z = 3 (large red triangles). There is generally good agreement between CAS selected mergers and statistically determined close pairs (with d $<$ 30 kpc) at all redshifts. Further, both methods agree that there is a rise in merger fraction with redshift, which is parameterised here by a simple power law best fit (solid line, with exponent = 3 +/-0.5) and power law exponential (dotted line), see Bluck et al. (2009) [Fig. 1] for further details on the fits. All displayed error bars are 1 $\sigma$ Poisson errors on the statistical counting.}
\end{figure}
\end{center}

\begin{figure*}[t!]
\begin{center}
\includegraphics[width=0.48\textwidth,height=0.48\textwidth]{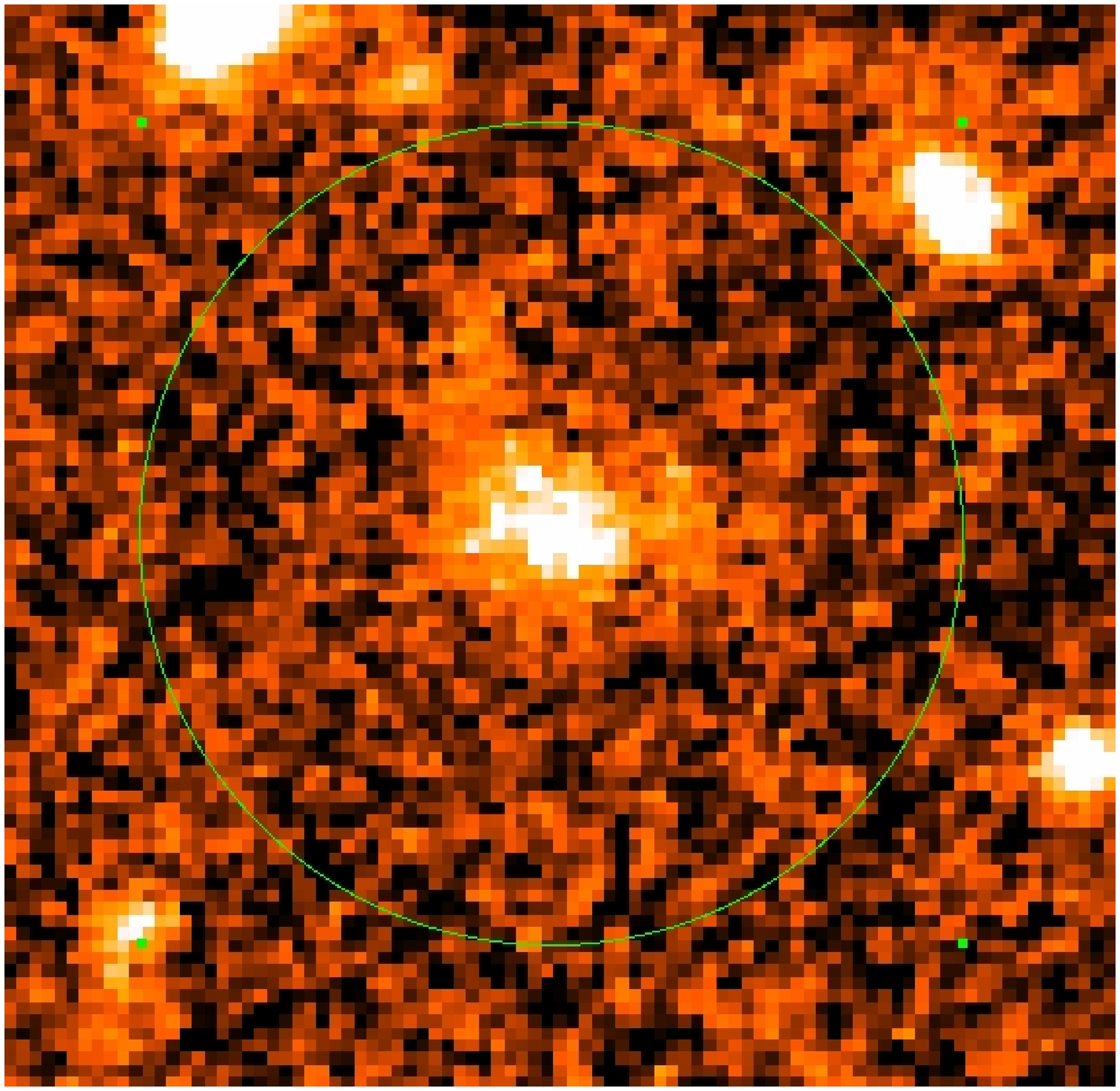}
\includegraphics[width=0.48\textwidth,height=0.48\textwidth]{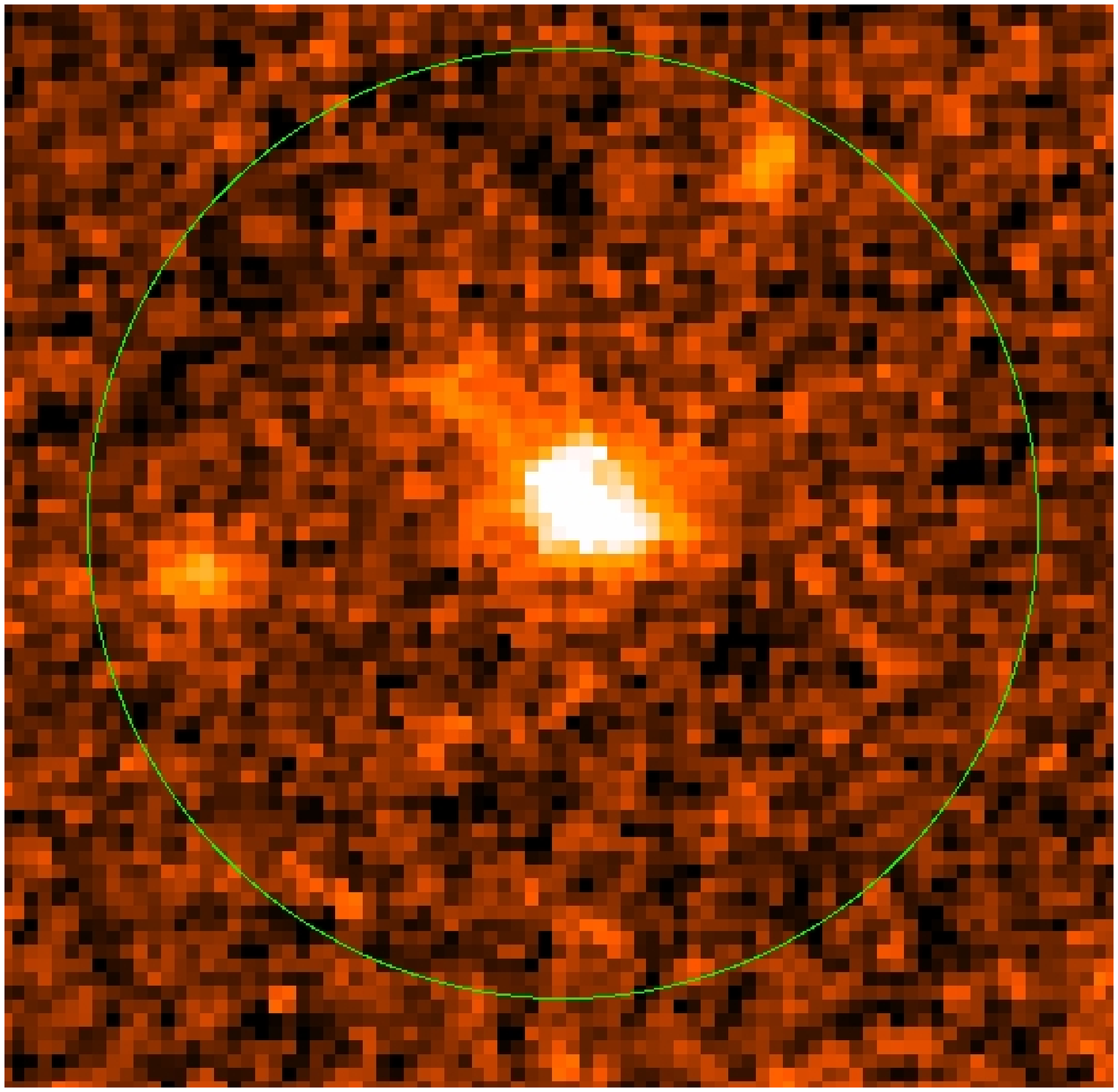}
\caption{Typical Examples of CAS Major Mergers: A $>$ 0.35. HST/NICMOS $H_{160}$ image of GNS-2282 (left) and GNS-3126 (right). The green circle indicates the d $<$ 30 kpc (in physical units) region we use to select close pairs within.}
\end{center}
\end{figure*}

\subsubsection{A link between Asymmetry (A) and GALFIT Residual Flux Fraction (RFF)}

The true advantage of non-parametric computational techniques in fitting galaxy stellar light profiles over parametric techniques (such as Sersic index fitting) is in the fact that no mathematical function is assumed to be appropriate for any given galaxy. In the cases of spheroidal or disk-like systems, fitting with Sersic indices can be hugely avantageous (see e.g. Peng et al. 2002, Trujillo et al. 2007, Buitrago et al. 2008), but for morphologically disturbed or irregular galaxies this is certainly not the case. It is probable that highly asymmetric galaxies (those with a high $A$) will be poorly fit by GALFIT and other procedures for determining a Sersic index. An excellent measure of the poorness of parametric fitting in the Sersic index case is the residual flux fraction (RFF, Blakeslee et al. 2006). Thus, we hypothesise that there might be some correlation between RFF and $A$. If this is so, we will have better demarcated the applicability of parametric and non-parametric techniques when applied to fitting galaxies. Moreover, this will provide additional evidence that a high fraction of our massive galaxies at high-z are in fact morphologically disturbed. 

Our sample of massive galaxies have previously had their light profiles fitted with a Sersic index approach using the GALFIT package (see Peng et al. 2002). Early results concerning the size evolution of massive galaxies and crude morphological divisions between disks and spheroids in our sample, as well as details on the fitting techniques used are presented in Buitrago et al. (2008). Specifically, the Sersic index and effective radii are obtained from fitting the following formula for the intensity of light at radii, $r$,

\begin{equation}
I_{r} = I_{e} \hspace{0.1cm} {\rm exp} \big( -b_{n} \big[ \big( \frac{r}{r_{e}} \big)^{1/n} - 1 \big] \big)
\end{equation}

\noindent where $I_{e}$ is the intensity of light at the effective radii, $n$ is the Sersic index (with a value of $n = 4$ being equivalent to de Vaucouleurs' (1948) formula for an elliptical galaxy, and a value of $n = 1$ giving rise to an exponentially declining surface brightness profile appropriate for a disk-like object). $r_{e}$ is the circularised effective radius, given by $r_{e} = a_{e} (1-\epsilon)^{1/2}$, where $\epsilon$ is the ellipticity and $a_{e}$ is the semi major axis of the best fit ellipse to the galaxy light profile at the effective radius, $r_{e}$. $b_{n}$ is a constant to be fit to the light profile.

The best fit Sersic profile is constructed via minimising the residuals, which are defined as the difference between the HST F160W (H) band image and the model fit. We quantify the residual flux fraction (RFF, see Blakeslee et al. 2006 and Hoyos et al. 2010), which is a proxy for `badness of fit', as:

\begin{equation}
{\rm RFF} = \frac{(\sum |Res| - 0.8 \times \sigma_{image})}{FLUX_{A_{2e}}}
\end{equation}

\noindent where the summation is performed over the total area within the ellipse fit of the galaxy within two effective radii, $A_{2e}$. $|Res|$ is the absolute (modulus) value of the residual image, and $FLUX_{A_{2e}}$ is the total flux within the the area of the ellipse within two effective radii for the observed image. $\sigma_{image}$ is defined as:

\begin{equation}
\sigma_{image} = \left(\sigma_{Bkg}^{2} + \frac{F}{g} \right)
\end{equation}

\noindent where $g$ is the effective gain of the NIC-3 camera, $F$ is the flux value from the model for each pixel, and $\sigma_{Bkg}$ is the standard deviation of the background (noise) from the observed H band images.

The RFF statistic computes the fraction of light, present or absent in the residual image, which is not explained by empirical errors in the Sersic model (for more details see Hoyos et al. 2010). This effectively equates to a measure of how reliable each fit is, with a lower number indicating a more reliable fit. Since this type of fitting assumes intrinsically symmetric light profiles of galaxies, it may be expected that it would not perform well on more asymmetric galaxies (or those with a high $A$ value, as defined in \S 3.1). To investigate the possibility of a correlation here between poorness of GALFIT fitting (RFF) and asymmetry ($A$) in high mass high redshift galaxies, we plot these variables against each other in Fig. 5.

We find that there is a correlation between RFF and $A$, whereby galaxies with higher assymetries tend to have worse fits from the GALFIT analysis. In fact we find that the mean value of $A$ rises with 3 $\sigma$ significance from low to high RFF values, taking the errors as standard error on the mean and incorporating into this the intrinsic errors from the RFF and $A$ methods. A best fit, least squares, simple power law to the unbanned data is found to be:

\begin{equation}
 A = (0.28\rm{+/-}0.13) \times (1 + RFF)^{1.6+/-0.7}
\end{equation}

\noindent This implies that asymmetry in the light profile of massive galaxies at high redshifts may be a significant impediment to the use of Sersic profile fitting for some systems. In these cases a non-parametric fit (such as CAS) will be more representative, since it makes no underlying assumptions about the spatial symmetry of the galaxy light profile. Furthermore, this relation opens the possibility to potentially identify likely candidates for mergers via high RFF values. For example, we find that galaxies with RFF $>$ 0.05 are $\sim$ 4 times more likely to be mergers (as defined by $CAS$, see \S 3.1.2) than those with RFF $<$ 0.05. From a KS test this represents a 3 $\sigma$ significant rise from low to high bins of RFF in $A$. There remains a high degree of scatter around the line of best fit. This is most probably explained by the errors inherent in measuring $CAS$ and the RFF, as very few galaxies lie further than 2 $\sigma$ from the best fit line. However, a much larger sample of galaxies with more general properties is needed to really establish this result. Nonetheless, our results are consistent with Sersic methods being successful for spheroids and disks, but failing with irregular galaxies, i.e. the dominant cause of poor GALFIT profile fitting is morphological disturbance for high mass high redshift systems. These results are also consistent with one of the overarching conclusions of this paper, i.e. that a significant fraction of high redshift massive galaxies are highly disturbed, most likely by major merging, with $\sim$ 30 \% of systems giving rise to a poor Sersic fit (RFF $>$ 0.05) and $\sim$ 10 \% leading to a very poor fit (RFF $>$ 0.1). This provides additional evidence that our sample of high-z massive galaxies are highly morphologically irregular to the non-parametric $CAS$ approach (outlined in the preceding sections): they are frequently not fit well by parametric approaches.

\begin{center}
\begin{figure}
\includegraphics[width=0.47\textwidth,height=0.47\textwidth]{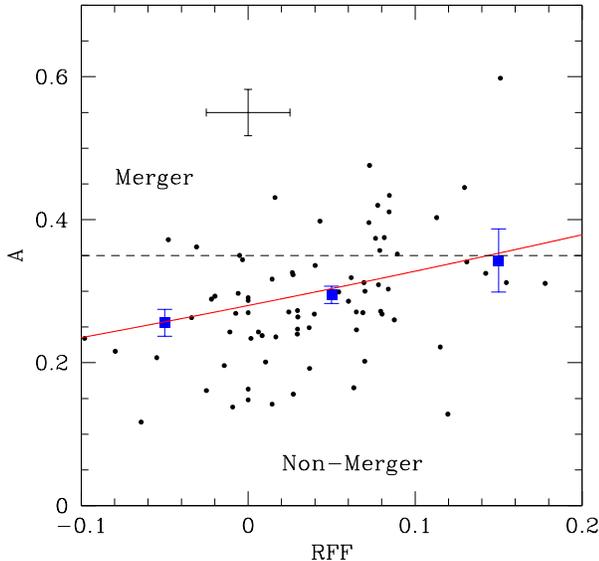} 
\caption{Asymmetry ($A$) vs. Residual Flux Fraction (RFF) from GALFIT Sersic index fitting, for our sample of massive ($M_{*} > 10^{11} M_{\odot}$) galaxies at z = 1.7 - 3. The black dots are the unbinned data, with the black error bar representing average 1 $\sigma$ errors on RFF and $A$. The blue squares represent mean values of A, in the RRF ranges -0.1 - 0, 0 - 0.1, and 0.1 - 0.2, with their corresponding error bars being 2 $\sigma$ errors on the mean in each case. The solid red line is a best fit simple power law to the unbinned data, which is defined in \S 3.1.3.}
\end{figure}
\end{center}

\subsection{Total Merger Fraction from Pair Statistics}

\subsubsection{Basic Method}

Most probes of the merger histories of massive galaxies are considerably limited by constraints due to the depth of the imaging used. This is especially the case at higher redshifts. For this reason the vast majority of merger histories in the literature (at least all those which extend beyond z $>$ 1) are restricted to `major' mergers. Although this definition varies from paper to paper, it is common to use a difference such as +/- 1.5 magnitudes ($>$ 1:4 by stellar mass) for close pairs, as in Bluck et al. (2009), which corresponds to a similar mass range for which $CAS$ is most sensitive (see also Conselice, Yang \& Bluck 2009 for a discussion of this). Some recent studies have probed deeper into the minor merger regime (e.g. Hopkins et al. 2010, Lopez-Sanjuan et al. 2011 and Lotz et al. 2011) but these usually only go down to a stellar mass ratio of 1:10 or less, and concentrate at lower redshifts than this study (z $<$ 1.5).

Due to the exceptionally deep HST NICMOS imaging we have acquired in the GNS, we are in a position to go much deeper, down to + 3.5 magnitudes at z = 3 and + 4.8 magnitudes at z = 2.3. This allows us to probe a mass range of up to a factor of $\sim$ 100 according to our photometric estimates, looking specifically at galaxies with masses $M_{*} = 10^{9} M_{\odot}$ - $10^{11.5} M_{\odot}$ around massive galaxies with $M_{*} > 10^{11} M_{\odot}$. Thus, we can probe the total merger history of the most massive galaxies in the Universe down to a threshold sensitivity of $M_{*} \sim 10^{9-9.5} M_{\odot}$, all the way back to z = 2 - 3. This considerably expands upon all previous attempts to ascertain the merger history of massive galaxies, by increasing the threshold sensitivity by over a factor of ten for high redshift galaxies. 

We define the total merger fraction, $f_{m}$, (in analogy to the major merger fraction of Bluck et al. 2009) to be:

\begin{equation}
f_{m} = \frac{1} {N} \sum_{i=1}^{i=N} (\rm{counts_{i} - corr_{i}})
\end{equation}

\noindent where

\begin{equation}
{\rm{corr}} = \int_{m-1.5}^{m+m^{+}} \rho \times \pi(r_{30kpc}^{2}-r_{5kpc}^{2}) dm'
\end{equation}

\noindent and $N$ is the total number of massive galaxies in the sample at each redshift range, $counts_{i}$ is the number of galaxies within the magnitude range $m - 1.5$ to $m + m^{+}$, contained within the annulus of 5 to 30 kpc (h = 0.7) around each massive host galaxy. We use $m - 1.5$ as the lower limit to our integral to assure that we count virtually all mergers down to the upper limit $m + m^{+}$. There are very few, if any, galaxies which are greater than 1.5 magnitudes brighter than the massive galaxies we probe at high redshifts, given that there are extremely few galaxies with $M_{*} > 10^{11.5} M_{\odot}$ at z $>$ 1.7.

The value $corr$ is the background correction, which we compute from the surface number density, $\rho$, of galaxies within the magnitude range under consideration, from our GNS survey. The fact that we take our surface number densities from the same survey (around the objects being studied at radii $<$ 200 kpc), which is centred around massive galaxies at high redshifts, reduces the risk of our results being unduly affected by differential redshift projection effects due to the increased clustering of massive galaxies at high redshifts (Bluck et al. 2009). This surface number density is then converted to an expectation value for the total number of close pairs one would measure by line of sight contamination, via multiplying over the area of sky contained within the annulus of 5 to 30 kpc around each massive host galaxy. We adopt this annulus method, as opposed to a simple circle approach, to avoid effects from the contamination of the light profiles of the merging galaxies by the host galaxy's brightness. Since these massive galaxies are extremely compact (see Buitrago et al. 2008) we find an inner radii of 5 kpc to be more than sufficient to remove this effect.

Typical contamination corrections vary accoding to the stellar mass ratio of the pairs, and the precise magnitudes of the galaxies in question. The major close pairs are corrected by typicaly up to a factor of two or so, with the more minor close pairs reduced by more than this, up to a factor of three to four. This is intuitive, because the broader the magnitude range used, hence the larger the mass ratio examined, the more contaminants from fore- and background galaxies arise in the sample. Furthermore, to convert from a pair fraction to a merger fraction requires use of a merger timescale. These are found to be in general longer for larger differences in the mass of two merging galaxies (see Lotz et al. 2008a,b and 2010 where this is determined through detailed N-body simulation); i.e. minor mergers take longer than major mergers, if all other factors, including the host, are the same. The exact form of the merger timescale dependence on mass ratio is unknown, however, but may lead to up to a factor of five difference in time from 1 : 4 major mergers to 1 : 100 minor mergers. All this leads to the frequency of minor merging being suppressed when compared to major merging, and not proportional to the galaxy stellar mass function, as one may naiively expect.

\subsubsection{Method Test - The Millennium Simulation}

The aim of this sub-section is to compare the merger fraction results obtained by the statistical close pair methods presented in this paper (and in Bluck et al. 2009) with spectroscopic close pair methods for a set of model galaxy catalogs constructed from the Millennium Simulation. In particular, we use light cone views developed by Kitzbichler \& White (2006) constructed from semi-analytic galaxy catalogs in the Millennium Simulation produced by De Lucia \& Blaizot (2007). The techniques used, and full details of the models are provided in these references. For the purposes of this section it should suffice to comment that the views were a set of six pencil beam light cone observations through an extended Millennium simulation, whereby 500 h$^{-3}$ Mpc$^{3}$ blocks were added together with a small angular shift to avoid exact repetition. The full prescriptions for assigning galaxy luminosities and masses from halo masses are beyond the scope of this paper, but are presented (for the interested reader) in detail in De Lucia \& Blaizot (2007) and references therein. Effectively the catalogs from De Lucia \& Blaizot (2007) provide the fundamental characteristics of the galaxies, such as stellar mass, luminosity, location in the model coordinates system, and the tables from Kitzbichler \& White (2006) provide the direct observational components, such as RA and Dec, observed (spectoscopic) redshifts and apparent K band magnetudes.

We selected eight areas of the Kitzbichler \& White (2006) views, with equal (0.01 square degree) areas which provide our eight unique runs. This was chosen to contain roughly equal high redshift massive galaxy numbers as our GNS sample, and to keep to a minimum the computational overheads that go with very large numbers of total galaxies visible in the view. The average number of $M_{*} > 10^{10.5} M_{\odot}$ galaxies at 1.7 $<$ z $<$ 3 in each run is 60 (comparable to the 80 used in the GNS sample). The average number of galaxies visible at all redshifts and stellar masses in each view is $\sim$ 32000. This provides ample numbers to gain meaningful statistics on the differences between spectroscopic and statistically deduced close pair fractions, from model data. We were forced to choose slightly lower stellar mass cuts than used in the GNS (log[$M_{*}$] $>$ 11) due to the intrinsic underpopulation of very massive galaxies (and halos) at high redshifts inherent in the Millennium simulation (see e.g. Bertone \& Conselice 2009 for a  discussion on this issue). However, the mass and redshift parameter space is broadly similar and should provide a good first formal test to the efficacy of the close pair method.

For each Millennium run we selected two samples: a high stellar mass, high redshift galaxy catalog containing only $M_{*} > 10^{10.5} M_{\odot}$ galaxies at 1.7 $<$ z $<$ 3; and a large catalog containing all galaxies viewed in the light cone and restricted area utilised, regardless of mass or redshift. We then ran our close pair codes (exactly as presented in \S 3.2.1 on our real data) and further ran an additional set of three spectroscopic close pair calculations (corresponding to massive galaxies with pairs within 30 kpc, with magnitudes +/-1.5 of the host galaxy magnitude, and velocity differences of $\Delta$v $<$ 500 km/s, 1000 km/s and 1500 km/s respectively) utilising the observed (spectroscopic) redshifts available for all objects in the view from the Kitzbichler \& White (2006) catalogs. Explicitly we compute the model spectroscopic pair fractions as:

\begin{equation}
f_{pair} = \frac{N_{pair}(<30{\rm kpc};^{+}_{-}1.5{\rm mag};\Delta V < X)}{N_{host}}
\end{equation}

\noindent Where $N_{pair}$ gives the total number of galaxies in the model with companions within 30kpc of the host, and within +/-1.5 of the host galaxy's K-band magnitude, with a velocity difference from the model observed spectroscopic redshift of X = 500 km/s, 1000 km/s and 1500 km/s. The statistical pair fraction is computed as in eq. (6) where the corrections for our statistical method are computed exactly as in eq. (7) in \S 3.2.1, with these corrections being taken around the objects in question (at radii $<$ 200 kpc in physical units). The results are presented in Table 2.

Graphs showing and comparing the modelled spectroscopically and statistically deduced pair fractions are presented in Fig. 6. We find that our statistical close pair method gives values which are in general lower than $\Delta$v $<$ 1500 km/s spectroscopic close pairs and higher than $\Delta$v $<$ 500 km/s. We also note that in all but one case (run 4) our statistical methods are slightly higher in value than the $\Delta$v $<$ 1000 km/s spectroscopic close pairs. Quantitatively, we calculate a mean difference of: +0.075+/-0.04 between statistical and spectroscopic close pairs (with $\Delta$v $<$ 500 km/s); +0.029+/-0.021 between statistical and spectroscopic close pairs (with $\Delta$v $<$ 1000 km/s); and -0.018 +/- 0.23 between statistical and spectroscopic close pairs (with $\Delta$v $<$ 1500 km/s). The errors presented here are taken as the standard deviation about the mean difference in values.

Care must be taken when utilising these raw results to attempt to transform these into updated errors on the statistical method. Since the environmental data on our sample of high redshift high stellar mass galaxies (presented in Gr\"utzbauch et al. 2011a,b) places these objects as primarily in groups, as oposed to field or cluster galaxies, this would seem to suggest that velocity dispersions ought to be $\sim$ 300 km/s. This is, unfortunately, not ideal when using the Millennium Simulation as the smoothing length inherent to the halo merger trees results in there being very few galaxies in close pairs with low ($\Delta$v $<$ 500 km/s) relative velocities for artificial reasons. As such, the exact velocities cannot be entirely trusted at small distances, which is the area of interest for this study. A full quantitative analysis of the effects of the smoothing length in the Millennium Simulation on over populating close pairs with higher velocities is beyond the scope of this paper. Nonetheless, it can be seen that the $\Delta$v $<$ 500 km/s spectroscopic close pairs scale less well with both the statistical method and the higher velocity difference spectroscopic methods, and we tentatively suggest that this is ultimately a result of the smoothing scale length in the Millennium Simulation. 

If we consider the $\Delta$v$<$ 1000 km/s spectroscopic close pair as the real values for close pairs (which is often used in the literature to incorporate galaxies with relative velocities $\Delta V < 3\sigma_{group}$), then we witness a relatively small offest of +0.029+/-0.021 from the modelled spectroscopic close pair fraction when compared to our statistical approach, which is smaller than the average statistical random errors for our high redshift data (+/- 0.07). It is also interesting to note that this effect is comfortably within 2 $\sigma$ of indicating no intrinsic bias at all. What is established here is that there is broad agreement between spectroscopic and statistical close pair methods when applied to the model data of the Millennium Simulation. There is a hint of a small possible bias for the statistical method to overestimate the spectroscopic method at $\Delta$v $<$ 1000 km/s, but this is only visible at the $<$ 2 $\sigma$ level and contributes a significantly smaller shift than the random errors on our data at high redshifts. This implies that we can reliably use our close pair method at high redshifts for massive galaxies, to an accuracy determined by the random errors.

\begin{center}
\begin{figure*}[t]
\includegraphics[width=0.48\textwidth,height=0.48\textwidth]{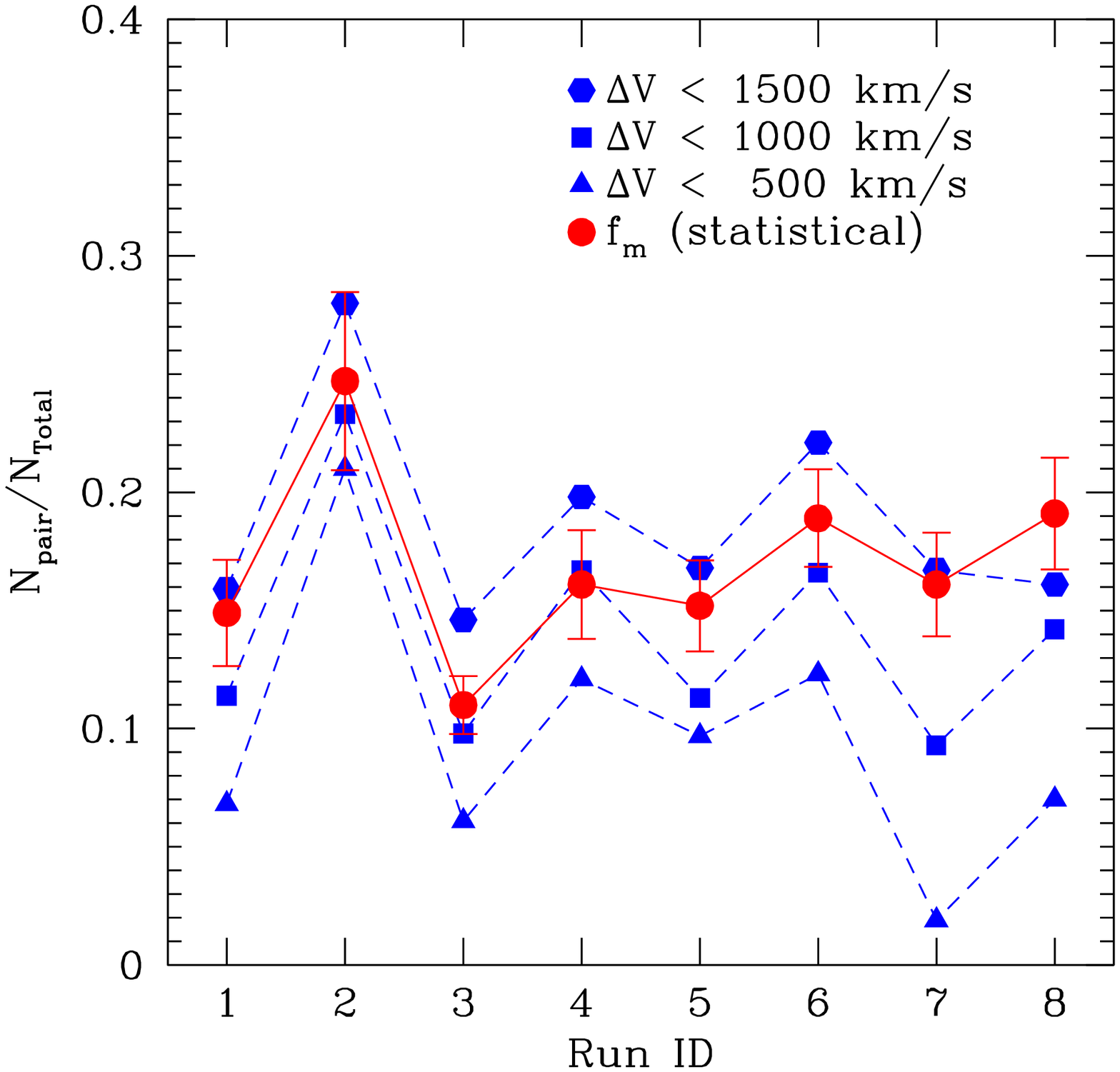}
\includegraphics[width=0.48\textwidth,height=0.48\textwidth]{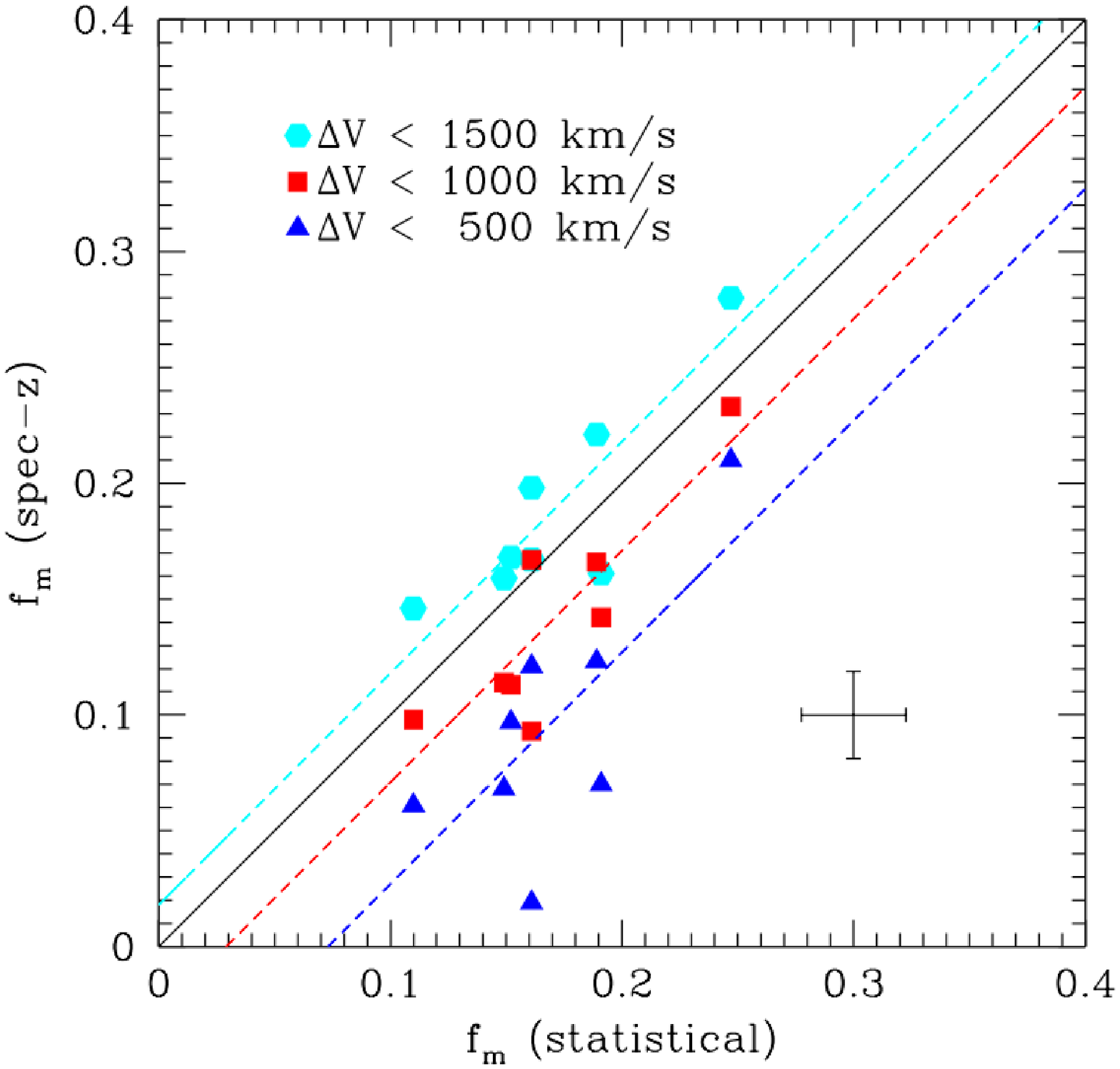}
\caption[Millennium Simulation]{Left panel is a comparison of modelled spectroscopic and statistical pair fractions ($f_{pair} = N_{pair}/N_{Total}$) across the eight runs performed on the Kitzbichler \& White (2006) views of the De Lucia \& Blaizot (2007) galaxy models drawn from the Millennium Simulation. The different shaped blue points are for modelled spectroscopic pair fractions using varying velocity differences, as shown on the plot. The red squares are the modelled statistical pair fractions of each independent run, with the errors on these being Poisson counting errors. These are the values we would measure using our method. Right panel presents model spectroscopically confirmed close pairs at varing velocity differences as a function of the modelled statistically deduced pair fraction (which we use on actual galaxies later in this paper). The black error bar represents the average counting error on all methods. These errors do not take into account cosmic variance, systematic effects or biases as this is in part what these plots aim to address. The solid black line indicates the one-to-one relation. The cyan, red and blue dashed lines indicate best fits to the $\Delta$v = 1500, 1000 and 500 km/s modelled spectroscopic close pair fractions respectively, assuming a constnt one-to-one gradient and varying the intercept only. Explicitly $f_{m}$ (spec-z) = $f_{m}$ (statistical) - c, where c = -0.018 for 1500 km/s, +0.029 for 1000 km/s, and +0.071 for 500 km/s.}
\end{figure*}
\end{center}

\subsubsection{Merger Timescales}

Unfortunately, the mechanics of the Millennium Simulation's semi-analytics makes it difficult to use for probing the merger timescales directly. There is neither sufficient resolution nor detailed enough gas physics implemented to trust the timescales between close pairs becomming mergers, thus, this is not explored further here. The most realistic models to date are implemented by Lotz et al. (2008a/b, 2010). These are based on gas rich disk mergers (Sbc models), and are computed via N-body simulations. The values they obtain are used towards the end of this paper, but it is acknowledged that uncertainty in the initial relative velocities remain an unrestricted source of error in these analyses at the present time.

Future planned spectroscopic study, as well as even more deatiled and realistic N-body simulations, will come to improve the constraints on this difficult issue in time. For now, adopting average group velocity dispersions of typically $\sigma_{group}$ $\sim$ 300 km/s and looking principally at gas rich disks have yielded workable values of the merger timescale ($\tau_{m}$). In light of the uncertainty in this issue, however, in the following sections the majority of the analyses are performed in a timescale independent manner, and where possible timescales are factored out of the equations until the final values are stated to make it easy for researchers in the future to implement new timescales into our analyses, when these are better constrained than at present.

\begin{figure*} [t]
\begin{center}
\includegraphics[width=0.48\textwidth,height=0.48\textwidth]{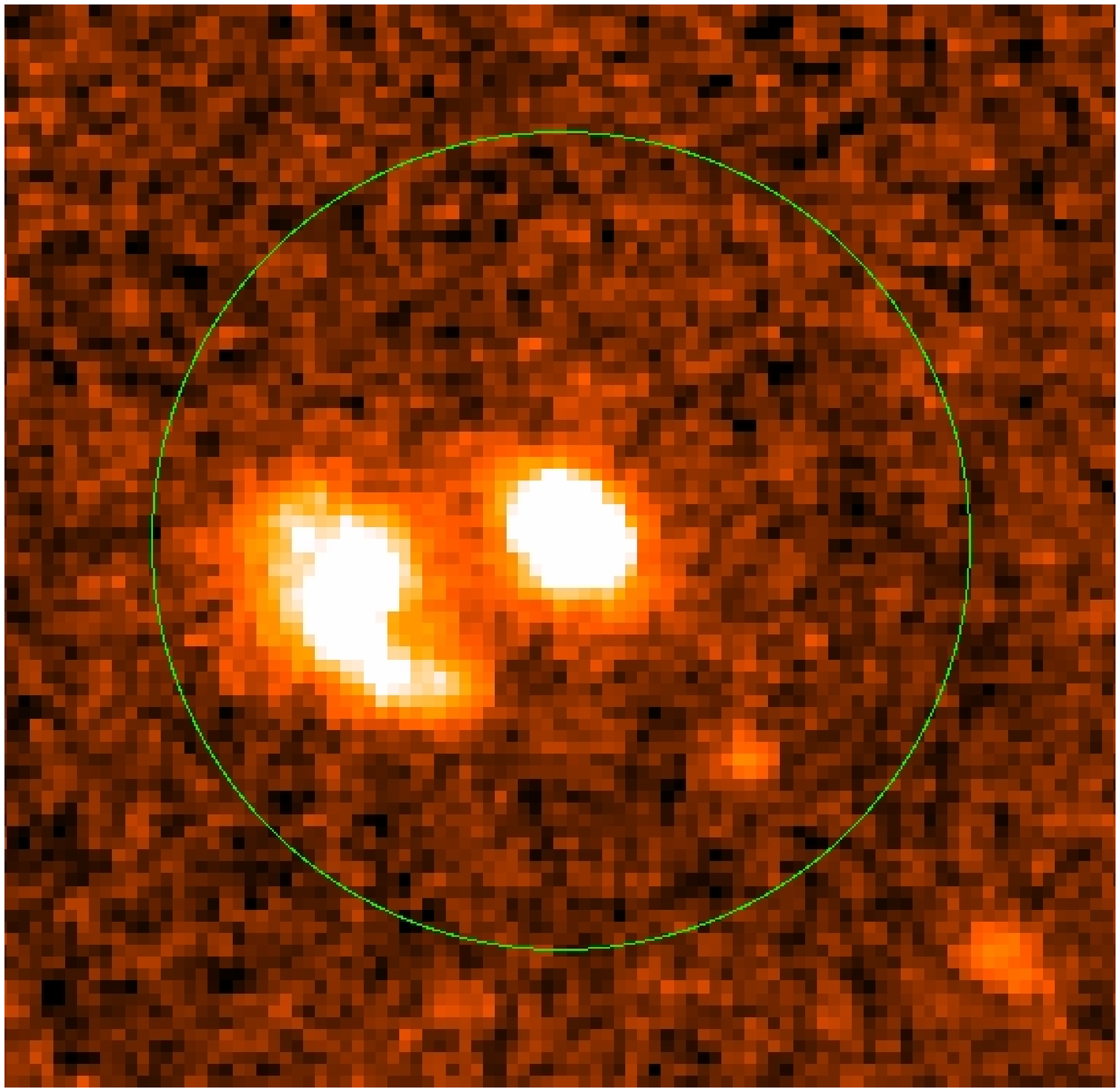}
\includegraphics[width=0.48\textwidth,height=0.48\textwidth]{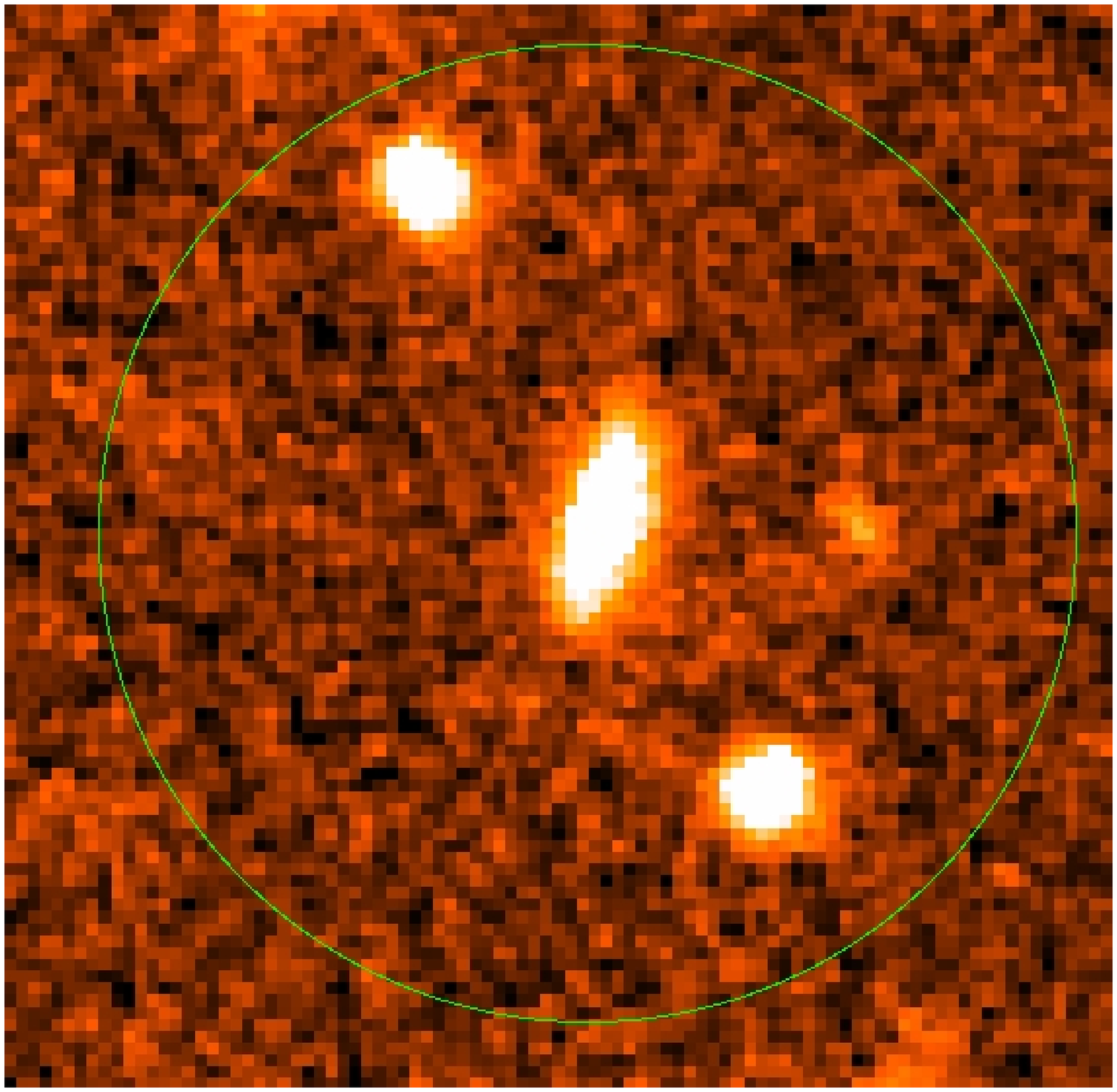}
\caption{Typical Examples of Major Close Pairs: 1:4 $< r_{M_{*}} <$ 4:1. HST/NICMOS $H_{160}$ image of GNS-8214 (left) and GNS-373 (right). The green circle indicates the d $<$ 30 kpc (in physical units) region we use to select close pairs within.}
\end{center}
\end{figure*}

\begin{figure*}
\begin{center}
\includegraphics[width=0.48\textwidth,height=0.48\textwidth]{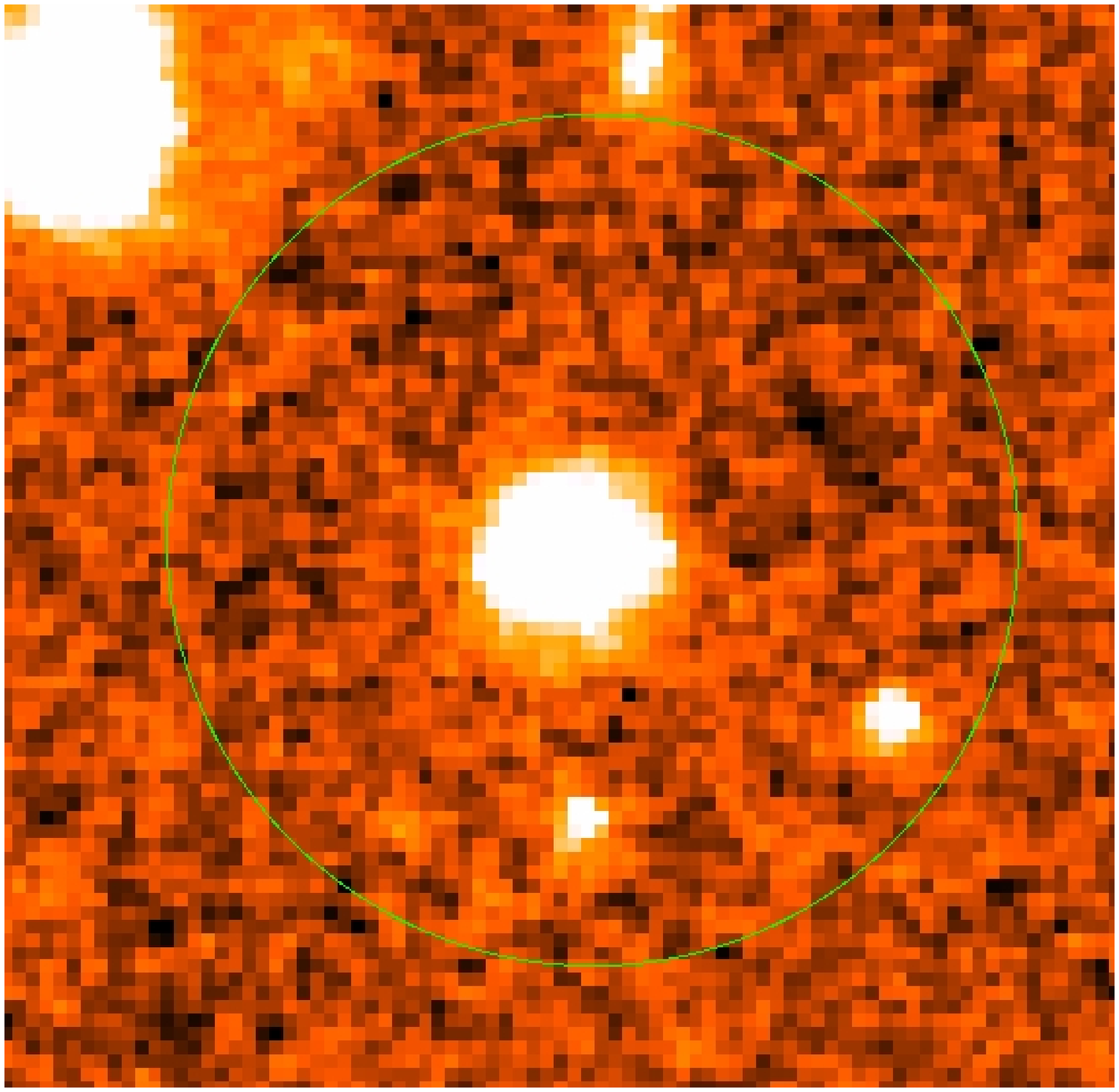}
\includegraphics[width=0.48\textwidth,height=0.48\textwidth]{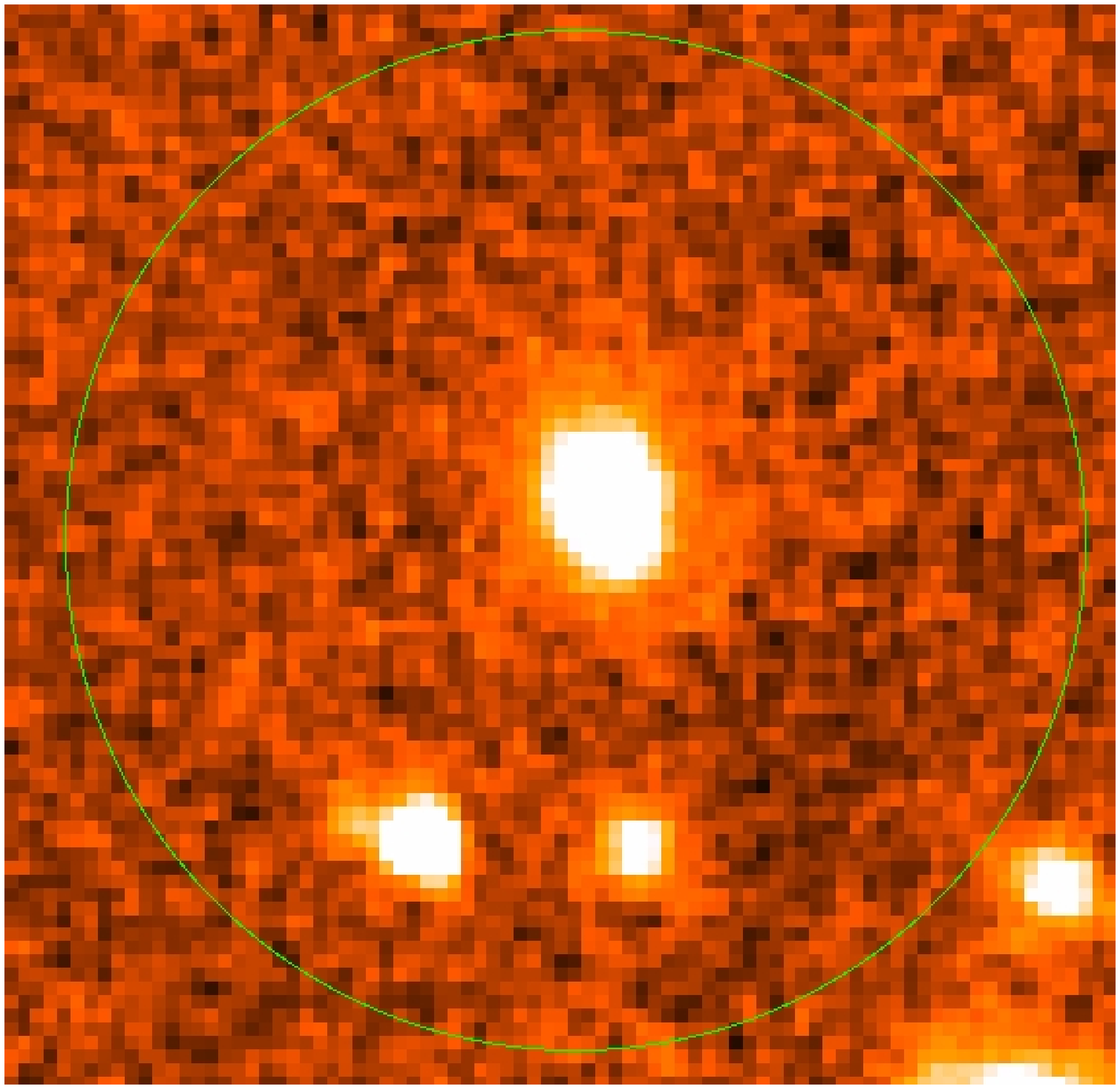}
\caption{Typical Examples of Minor Close Pairs: 1:100 $< r_{M_{*}} <$ 1:4. HST/NICMOS $H_{160}$ image of GNS-7979 (left) and GNS-840 (right). The green circle indicates the d $<$ 30 kpc (in physical units) region we use to select close pairs within.}
\end{center}
\end{figure*}

\subsubsection{Total Merger History: Results and Fitting}

In order to probe the total merger history of massive galaxies at high redshifts, we vary the upper limit to the magnitude range of interest ($m^{+}$ in equation 7) from 1.5 to 6.5 in increments of 1.5 (covering a mass ratio up to $M_{*}^{host} / M_{*}^{pair}$ $\sim$ 100), probing the total merger fraction. We also probe an ultra-major merger fraction ($>$ 1:2 by mass) of m$^{+}$ = 0.5. See Fig. 7 for typical examples of major close pairs, and Fig. 8 for typical examples of minor close pairs. The results of this are plotted in Fig. 9. The vertical dotted lines indicate the completion limits for each redshift range probed from the GNS (see Conselice et al. 2011a for full details on deducing our mass and magnitude completion thresholds). Essentially we compare our number counts of galaxies at differing magnitudes to those of the much deeper HDF, and conclude that our limits are where these curves cease to agree closely (data for the HDF-N comes from Thompson et al. 1999 and for the HDF-S from Metcalfe et al. 2006).

We witness a rise in merger fraction with magnitude range out to the completion limits. We note that the major and minor pair systems are in general different populations, with only $\sim$ 20 \% of major mergers also being identified with minor merging. We also present a rough guide to the masses of objects included, at the top of Fig. 9. This is computed by assuming a mass-to-light ratio of 2.5 : 1 (i.e. $\Delta H_{AB} / \Delta log M_{*}$ = 2.5) as frequently implemented in the literature, and a mean stellar mass for the host galaxies of $M_{*} \sim 10^{11} M_{\odot}$.

We fit the merger fraction dependence on magnitude range, up to the respective completion limits (of + 3.5 mag at z = 3 and + 4.8 mag at z = 2.3). At 2.3 $<$ z $<$ 3 we find the merger fraction dependence on magnitude range is:

\begin{equation}
f_{m}(\delta H_{AB}) = (0.12\rm{+/-}0.07) \times (1.5 + \delta H_{AB})^{0.91+/-0.35} ;
\end{equation}

\noindent at 1.7 $<$ z $<$ 2.3 we find a slightly steeper dependence of:

\begin{equation}
f_{m}(\delta H_{AB}) = (0.04\rm{+/-}0.03) \times (1.5 + \delta H_{AB})^{1.24+/-0.37} .
\end{equation}

\noindent Assuming a mass-to-light ratio, the above equations may be converted to the merger fraction dependence on mass range ($\delta m_{*}$). We use a mass-to-light ratio of 2.5 : 1 (L : M), and define $\delta m_{*} = 11.5 - \rm{log}(M_{*}^{pair}[M_{\odot}])$, where $M_{*}^{pair}$ is the stellar mass of the paired galaxy, and log($M_{*}$) = 11.5 is chosen as the upper limit to a major merger, since there are extremely few, or no, galaxies with stellar masses greater than this at z $>$ 1.7. Thus, at 2.3 $<$ z $<$ 3 we find a merger fraction dependence on mass range for massive galaxies of:

\begin{equation}
f_{m}(M_{*}) = (0.28\rm{+/-}0.17) \times \delta m_{*}^{0.91+/-0.35}
\end{equation}

\noindent and at 1.7 $<$ z $<$ 2.3 we find a merger fraction dependence on mass range for massive galaxies of:

\begin{equation}
f_{m}(M_{*}) = (0.12\rm{+/-}0.09) \times \delta m_{*}^{1.24+/-0.37} .
\end{equation}

\noindent This gives a prescription for deducing the probable total (minor + major) number of mergers from extrapolation of these trends, via the following methodology. 

The total number of mergers, within a given mass and redshift range, is given by calculating the double integral:

\begin{equation}
N_{m} = \int_{M_{1}}^{M_{2}} \int_{t_{1}}^{t_{2}} \frac{1}{\Gamma(z,M_{*})} \hspace{0.1cm} dt \hspace{0.2cm} dM_{*}
\end{equation}

\begin{equation}
= \int_{M_{1}}^{M_{2}} \int_{z_{1}}^{z_{2}} \frac{1}{\Gamma(z,M_{*})} \frac{t_{H}}{(1+z)} \frac {dz}{E(z)} \hspace{0.2cm} dM_{*}
\end{equation}

\noindent where,

\begin{equation}
\Gamma(z, M_{*}) = \frac{\tau_{m}(M_{*})} {f_{gm}(z, M_{*})}
\end{equation} 

\noindent and

\begin{equation}
f_{gm}(z,M_{*}) = \frac{2 \times f_{m}(z, M_{*})} {1 + f_{m}(z,M_{*})} .
\end{equation}

\noindent $\Gamma(z,M_{*})$ is the characteristic time between mergers, $t_{H}$ is the Hubble time, $f_{m}(z,M_{*})$ is the merger fraction, $f_{gm}(z)$ is the galaxy merger fraction (i.e. the number of galaxies merging, not the total number of mergers), $\tau_{m}(M_{*})$ is the merger timescale (which we assume to be redshift invariant), and the parameter $E(z) = [\Omega_{M}(1+z)^{3}+\Omega_{k}(1+z)^2+\Omega_{\Lambda}]^{1/2} = H^{-1}(z)$, which we evaluate for a flat spacetime with a $\Lambda$CDM cosmology as defined in \S 1.

We construct a parameterisation of the total galaxy merger fraction dependence on redshift, derived at our mass completion threshold of log($M_{*}$) $\sim$ 9, thus:

\begin{equation}
f_{gm}(z) |_{M_{*} > 10^{9} M_{\odot}} = f_{gm}(z=0) |_{M_{*} > 10^{9} M_{\odot}} \times (1+z)^{\alpha} .
\end{equation}

\noindent Since we have only two independent redshift values for the total galaxy merger fraction derived at our completion limit, we are not particularly well able to constrain the form of this function at present. As such, to model our errors, we allow the exponent, $\alpha$, to vary from 0.5 - 5 and compute least squares best fits to $f_{gm}(z=0)$, taking into account the errors on our initial data points. We find that $f_{gm}(z=0)$ varies from 0.40 at $\alpha$ = 0.5 to 0.001 at $\alpha$ = 5. This range in $\alpha$ is wide enough to incorporate almost all measured values of the merger fraction dependence on redshift for major and minor mergers computed to date (e.g. Patton et al. 2000, Rawat et al. 2008, Conselice et al. 2008, Bluck et al. 2009, Conselice, Yang \& Bluck 2009, Lopez-Sanjuan et al. 2010a/b, Hopkins et al. 2010).  We then use this variance as an estimator to compute our underlying error in measuring the total number of mergers.

To evaluate the total number of mergers a massive galaxy will experience in a given redshift interval we must simplify equations 13 and 14, and apply these to our mass range of interest. Explicitly we calculate the total number of mergers through this summation, an approximate numerical solution to the integration in equation 14, evaluated for our mass range of interest:

\begin{equation}
N_{m} \approx \frac {1} {<\tau_{m}(M_{*})>} \sum_{z=z_{1}}^{z=z_{2}} \left(  \frac{t_{H} f_{gm}(z) |_{M_{*} > 10^{9} M_{\odot}}}{(1+z)E(z)} \right) \delta z
\end{equation}

where,

\begin{equation}
<\tau_{m}(M_{*})> = \frac {\int_{M_{1}}^{M_{2}} \tau_{m}(M_{*}) \hspace{0.15cm} dM_{*} } {\int_{M_{1}}^{M_{2}} \hspace{0.15cm} dM_{*} } .
\end{equation}

\noindent $M_{1} = 10^{9} M_{\odot}$ and $M_{2} = 10^{11.5} M_{\odot}$ for our study. All other variables are as defined above for equations 13 - 17. Unfortunately, since the form of the function $\tau_{m}(M_{*})$ is currently not well constrained we cannot compute the mean merger timescale directly here, but we include the general form for completion.

Evaluating this summation from $z_{1}$ = 3 to $z_{2}$ = 1.7 (the area probed directly in this study), we find that an average massive ($M_{*} > 10^{11} M_{\odot}$) galaxy will experience on average $N_{m}$ = (1.1 +/- 0.2) / $<\tau_{m}(M_{*})>$[Gyrs]) mergers with galaxies with $M_{*} > 10^{9} M_{\odot}$ over the redshift range z = 1.7 - 3. We estimate a total of $N_{m}$ = (4.5 +/- 2.9) / $<\tau_{m}(M_{*})>$[Gyrs] mergers with $M_{*} > 10^{9} M_{\odot}$ galaxies from z = 3 to the present, via extrapolation of the trend. Note that the errors are considerably larger here due to our ignorance of the exponent of the redshift dependence of the galaxy merger fraction ($f_{gm}$), and hence the characteristic time between mergers ($\Gamma$) used in the integral computing the total merger number, $N_{m}$. We factor the merger timescale $\tau_{m}(M_{*})$ out of our calculations (assuming that it is redshift independent) as this is poorly known for the mass range of galaxies probed in this paper. However, as a rough guide, Lotz et al. (2008a,b and 2010) suggest that this might be $\sim$ 0.4 Gyr for 3:1 mass mergers, and $\sim$ 1 Gyr for 9:1 mass mergers from N-body simulations. This appears to be roughly a logarithmic dependence of timescale on stellar mass, suggesting that at 1:100 $\tau_{m}$ $\sim$ 2 - 3 Gyrs, but this is currently unconstrained directly either by N-body simulations or observational approaches.

In order to calculate the stellar mass increase due to these major and minor mergers, we must compute an average mass increase per merger. This is computed via the weighted mean from fits to the merger fraction dependence on mass range (see Fig. 9 and eqs. 11 and 12). Specifically, we compute:

\begin{equation}
\frac {< \Delta M_{*} >} {N_{m}} = \int_{z_{1}}^{z_{2}} \left( \frac{\int_{M_{1}}^{M_{2}} M_{*} f_{gm}(z,M_{*}) dM_{*}} {\int_{M_{1}}^{M_{2}} f_{gm}(z,M_{*}) dM_{*}} \right) \frac{dz}{z}
\end{equation}

\noindent where we integrate over our mass range of interest from $10^{9} - 10^{11.5} M_{\odot}$. We note that the form of the function $f_{gm}$, i.e. its exponent, does not vary strongly with redhsift between z = 1.7 and 3 (the range we probe directly). So, we take its average value and compute the errors from the extremes allowed by the fits. If, however, this is found to vary strongly with redshift from z = 1.7 to the present, our extended results will need to be recomputed accordingly in the future when deeper and wider low redshift data is available.

We find $< \Delta M_{*} > / N_{m} = (7.5 +/- 2.6) \times 10^{10} M_{\odot}$. In the redshift range probed directly by the GNS (z =  1.7 - 3), this corresponds to an average massive galaxy experiencing a mass increase of $< \Delta M_{*} > = (8.3 +/- 3.2) / <\tau_{m}(M_{*})>[\rm{Gyrs}] \times 10^{10} M_{\odot}$. This leads to a total stellar mass increase over the past $\sim$ 12 Gyrs (z = 3 - 0) from extrapolation, due to $M_{*} > 10^{9} M_{\odot}$ mergers, with massive ($M_{*} > 10^{11} M_{\odot}$) galaxies, of $< \Delta M_{*} > = (3.4 +/- 2.2) / <\tau_{m}(M_{*})>[\rm{Gyrs}] \times 10^{11} M_{\odot}$. 

A good first order approximation to the actual value of $\tau_{m}(M_{*})$ will be to take its value at our median mass of merger of $\sim 2 \times 10^{10} M_{\odot}$, i.e $\sim$ 1 Gyr from N-body simulations in Lotz et al. (2010).  When compared to the major merger stellar mass increase, computed in Bluck et al. (2009), of $(1.7 +/- 0.5) \times 10^{11} M_{\odot}$, this suggests that minor mergers contribute a similar amount of stellar mass to major mergers from z = 3 to the present, with a minor merger stellar mass increase of $\sim 1.7 \times 10^{11} M_{\odot}$. It is still, however, possible that ultra-minor mergers of massive galaxies with $M_{*} < 10^{9} M_{\odot}$ galaxies will be so numerous that they will significantly effect the mass increase of massive galaxies, although this seems very unlikely given the trends and analyses presented in this paper, and the longer timescales they will experience due to larger differences in mass ratio.

The average stellar mass of our GNS massive galaxy sample is $< M_{*} > = 1.7 \times 10^{11} M_{\odot}$. Thus, the mass increase due to all mergers down to a stellar mass sensitivity threshold of 1:100 over the redshift range probed directly by this study ( z = 3 to z = 1.7) is $\delta M_{*} / M_{*}$ = 0.49 +/- 0.21. With $\sim$ 65 \% of this increase coming from major mergers and $\sim$ 35 \% being due to minor mergers. Further extrapolation of this trend to z = 0 suggests that there may be a total mass increase $\delta M_{*} / M_{*}$ = 2.0 +/- 1.2 per galaxy, with roughly equal contribution from major and minor mergers over this total period. This indicates a factor of three growth in mass, i.e. $M_{*}^{final} = (1 + \delta M_{*} / M_{*}^{initial}) M_{*}^{initial} = 3 \times M_{*}^{initial}$. But this value should be taken with more caution as we are unable to probe directly the evolution from z = 1.7 to the present. However, this value is consistent with other recent work on the minor merger histories of massive galaxies at lower redshifts in Lopez-Sanjuan et al. (2011) where they find a $\delta M_{*} / M_{*}$ $\sim$ 0.4 down to a stellar mass ratio of 1:10 from z = 1 to the present, using a maximum likelihood close pair method. In the same stellar mass and redshift regime, we find $\delta M_{*} / M_{*}$ $\sim$ 0.3 +/- 0.25 (note that our errors are large due to us not probing this range directly in our study). Nonetheless, this demonstrates a consistency and possible complementarity between the approaches used here and in the work of Lopez-Sanjuan et al. (2011).

\hspace{3cm}

\section{Discussion}
\subsection{Mergers}
In this paper we have utilised several different methods to probe the structures and evolution of massive galaxies over cosmic time. First, it is important to acknowledge that there are extremely massive galaxies at z $>$ 2, but their properties are somewhat different to local massive galaxies (see e.g. Mortlock et al. 2011 for a study of the massfunction at these redshifts). The most prominent difference is that these massive galaxies tend to have smaller effective radii for their masses (e.g. Buitrago et al. 2008). We find that $\sim$ 1/4 of all massive galaxies (with $M_{*} > 10^{11} M_{\odot}$) are highly morphologically disturbed at 1.7 $<$ z $<$ 3, and in fact fit the rest frame optical definition for a CAS selected merger (see \S 3.1). This implies that major merging is an important factor in the evolution of massive galaxies, and qualitatively supports a hierarchical approach to galaxy formation. Moreover, we find close accord between (d $<$ 30 kpc) close pair fractions and CAS merger fractions at all redshifts, up to z = 3 (see Fig. 3). This strongly suggests that both approaches trace the underlying major merger activity in massive galaxies, and, furthermore, implies that the timescales involved in each approach must be similar (see Conselice, Yang \& Bluck 2009 for a discussion of this effect at lower redshifts). 

By way of comparison to other techniques, the stellar mass increase we compute from the total merger history down to a stellar mass threshold of $M_{*} > 10^{9} M_{\odot}$ agrees favourably with studies of the evolution of the mass function of massive galaxies from z = 3 to the present, see e.g. Mortlock et al. (2011). By examining the stellar mass function's evolution over cosmic time, one can place an approximate upper limit to the mass increase massive ($M_{*} > 10^{11} M_{\odot}$) galaxies can undergo from z = 3 to the present, of less than a factor of eight. This value, however, has large errors of approximately a factor of two or so on the upper limit. Further analysis of star formation in these massive galaxies (performed in Bauer et al. 2011) lend an additional factor of two or so stellar mass increase via star formation. Therefore, by merging and star formation we could expect our massive galaxies to increase in mass by around a factor of four to six, consistent with the less than a factor of eight prediction from analysis of the stellar mass function evolution. Ultimately, this suggests that both merging (major and minor) and star formation are important in the stellar mass build up of massive galaxies over the past 12 Gyr. The final piece to this puzzle is cool gas accretion onto galaxies, which will be considered in a forthcoming paper in this GNS series (Conselice et al., in prep.).

\begin{figure} 
\includegraphics[width=0.47\textwidth,height=0.47\textwidth]{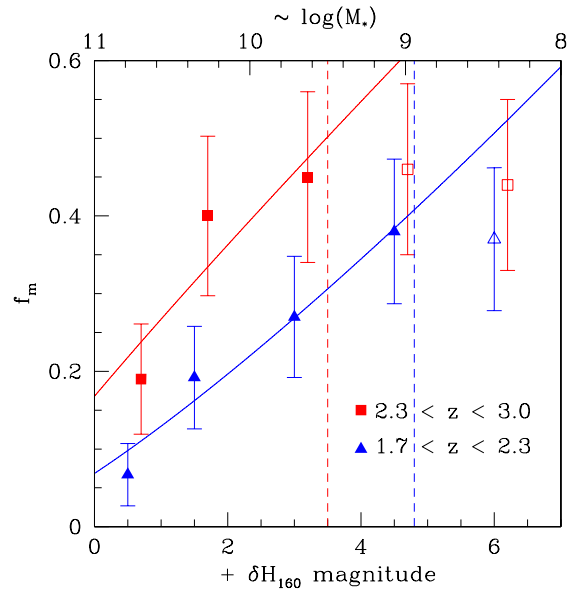} 
\caption{The Merger Fraction ($f_{m}$) as a function of magnitude range (+$\delta H_{160}$) and approximate stellar mass sensitivity threshold ($\sim$ log($M_{*}$)) for massive galaxies (with $M_{*} > 10^{11} M_{\odot}$). The data points are separated into two redshift ranges as displayed on the plot. The error bars displayed are 1 $\sigma$ Poisson errors on the statistical counting. The vertical dotted lines indicate the completeness limits for each redshift range, respectively. Solid symbols indicate merger fractions computed where we are complete, and open symbols indicate merger fractions computed where we are incomplete. The solid lines are best fit simple power law functions to the merger fraction dependence on mass/ magnitude range for each redshift range (see \S 3.2.4 for further details).}
\end{figure}

\subsection {Size Evolution}

Numerous studies over the past five years have found increasingly compelling evidence for dramatic size evolution of massive galaxies over the past ten or so billion years (e.g. Trujillo et al. 2007, Buitrago et al. 2008, Cimatti et al. 2008, van Dokkum et al. 2009, Carasco et al. 2010). It is currently widely argued that massive galaxies may grow in size by up to a factor of five since z = 3, due largely to stellar mass build up in the outer regions (or wings) of these galaxies. Whilst there are still concerns that some of this apparent size evolution may be driven by Sersic fitting of massive galaxies at high redshifts not being the best way to represent the luminosity profiles (e.g. Mancini et al. 2010), a wealth of evidence from simulation, and observational attempts to break the degeneracies inherent in former approaches, are coming to establish the validity of this result (e.g. Capellari et al. 2009, Onodera et al. 2010). If this size evolution is real, then it becomes of vital importance in contemporary astrophysics to address the question: {\it What drives the size increase of massive galaxies over cosmic time?}

Naab et al. (2009) and Khochfar \& Silk (2006) have both argued from a theoretical standpoint for merger driven size evolution of massive galaxies. We have, in this paper, estimated the total merger history of massive galaxies with $M_{*} > 10^{9} M_{\odot}$ galaxies (see \S 3.2.4). We found that there are $\sim$ 4.5 mergers predicted from z = 3 to the present, which may go some way to explaining the observed size evolution of massive galaxies. In fact, arguments presented in Naab et al. (2009) suggest that a mass increase due to minor and major mergers of a factor of two, could lead to size evolution of up to a factor four in radii. The more mergers there are the greater the size evolution, thus a mass increase of a given magnitude will be more effective at increasing the size of the resultant galaxy if it is imparted in a series of minor mergers rather than one major merger. Thus, it is the relatively high fraction of mass imparted by minor mergers which is most useful for increasing a massive galaxy's size. 

This can be seen quantitatively by assuming a virialised merging system, such that the final merged galaxy is virialised, and the two merging galaxies are also virialised. This is a realistic and theoretically motivated scenario as Khochfar \& Burkert (2006) show that dark matter halos merge predominantly on parabolic orbits, thus conserving energy, within current leading cosmological simulation. Thus, from the virial formula, 2K + W = 0, we have (as in Naab et al. 2009, Binny \& Tremaine 2008):

\begin{equation}
E_{i} = K_{i} + W_{i} = -K_{i} = \frac{1}{2} W_{i}
\end{equation}
\begin{equation}
= - \frac{1}{2} M <v_{i}^{2}> = -\frac{1}{2} \frac{GM_{i}^{2}}{r_{g,i}}
\end{equation}

\noindent Where $E_{i}$ is the total energy, $M_{i}$ is the total mass, $<$$v_{i}^{2}$$>$ is the mean square velocity, and $r_{g,i}$ is the gravitational radius of the initial host massive galaxy (i.e. the radius at which mass $M_{i}$ is contained). By defining the total energy of the (major or minor) merging galaxy (as in Naab et al 2009) as $E_{a}$, and the total energy of the final resultant merged galaxy as $E_{f}$, with all other variables similarly labelled (e.g. $M_{a}$ is the mass of the major or minor merging galaxy etc.), we have, assuming a fully virialised system:

\begin{equation}
E_{f} = - \frac{1}{2} \frac {GM_{f}^2}{r_{g,f}} =  E_{i} + E_{a} = - \frac{1}{2} \frac {GM_{i}^2}{r_{g,i}} - \frac{1}{2} \frac {GM_{a}^2}{r_{g,a}}
\end{equation}

\begin{equation}
= - \frac{1}{2} \frac{(1 + \eta \epsilon) GM_{i}^2 }{r_{g,i}} = - \frac{1}{2} \frac{(1 + \eta)^{2} GM_{i}^2}{r_{g,f}} .
\end{equation}

\noindent Therefore,

\begin{equation}
\frac{r_{g,f}}{r_{g,i}} = \frac{(1 + \eta)^{2}}{(1 + \eta \epsilon)} ;
\end{equation}

\noindent where,

\begin{equation}
\eta = M_{a} / M_{i} \hspace{0.5cm} {\rm and} \hspace{0.5cm} \epsilon = <v_{a}^{2}> / <v_{i}^{2}> .
\end{equation}

\noindent See Naab et al. (2009) for the original (slightly different) derivation and a detailed discussion.

The result of this is that minor mergers are a more efficient driver of galaxy growth than major mergers. This is because, for a similar total mass increase, there will be a greater growth in size if this mass is imparted as a series of smaller mergers rather than one big merger due to the smaller value of $\epsilon$ in this case. Further details on these arguments as well as detailed derivations can be found in Naab et al. (2009), Hopkins et al. (2009) and Bezanson et al. (2009).

For the mass increase due to minor mergers witnessed in this paper, of roughly a factor of 2 from z = 3 to the present, corresponding to $\eta_{minor} \sim 1$ and $\epsilon_{minor}$ $<<$ 1, there should be a size increase of up to a factor of four. Furthermore, the mass increase due to major merging of order another factor of two by mass, corresponding to $\eta_{major} \sim 1$ and $\epsilon_{major} \sim 1$, would lead to a growth in size of up to a further factor of two. These estimates should be taken to be upper limits as it is probable that energy will be lost to the intergalactic medium during merging, via a number of processes, including the formation of new stars, heating of interstellar gas, and igniting of AGN. Therefore, we can place an upper limit on the amount of massive galaxy growth allowable via major and minor merging from z = 3 to the present of a factor of $<$ 6. This is comparable to the observed factor of $\sim$ 4 - 5 growth in effective radii (seen in e.g. Buitrago et al. 2008) suggesting that merging can be primarily responsible, with minor merging being the more efficient route, but that there may need to be additional contributors as well since these are upper limits. See Fig. 10 for the allowable mass and size evolution for massive galxies from z = 3 to the present.

This result is broadly consistent with recent work from Trujillo et al. (2011) where the authors conclude that $\sim$ 8 minor (1:10) mergers are required, or else $\sim$ 3 major (1:3) mergers, to bring high redshift compact massive galaxies to the appropriate local mass - radius relation. We find that $\sim$ 4 - 5 mergers (with a range of stellar mass ratios from 2 : 1 to 1 : 100) takes a massive galaxy between 50 - 100 \% of the required distance to the local relation from its average z = 2 - 3 location.

Therefore, the minor and major merger history witnessed in this work may be large enough to explain most of the growth witnessed in massive galaxies across the past 12 billion years (e.g. Trujillo et al. 2007, Buitrago et al. 2008) according to recent theoretical estimation (Naab et al. 2009). It is, however, still worth considering other contributive factors, such as the effect of the enormous outpouring of energy from supermassive black holes in the nuclei of massive galaxies over this same period of cosmic history (see Fan et al. 2008 and Bluck et al. 2011). It may yet prove probable that it is a combination of star formation, passive evolution, major and minor merging, and AGN `puffing up' which causes the size evolution of massive galaxies. This work, however, implies that merging by itself could be responsible for most of the observed size evolution, assuming the validity of the arguments presented in Naab et al. (2009) and appropriate gas fractions (dictating the energy losses complicating the purely virialised merger model), which are yet to be empirically deduced.

\begin{figure*}
\begin{center}
\includegraphics[width=0.48\textwidth,height=0.48\textwidth]{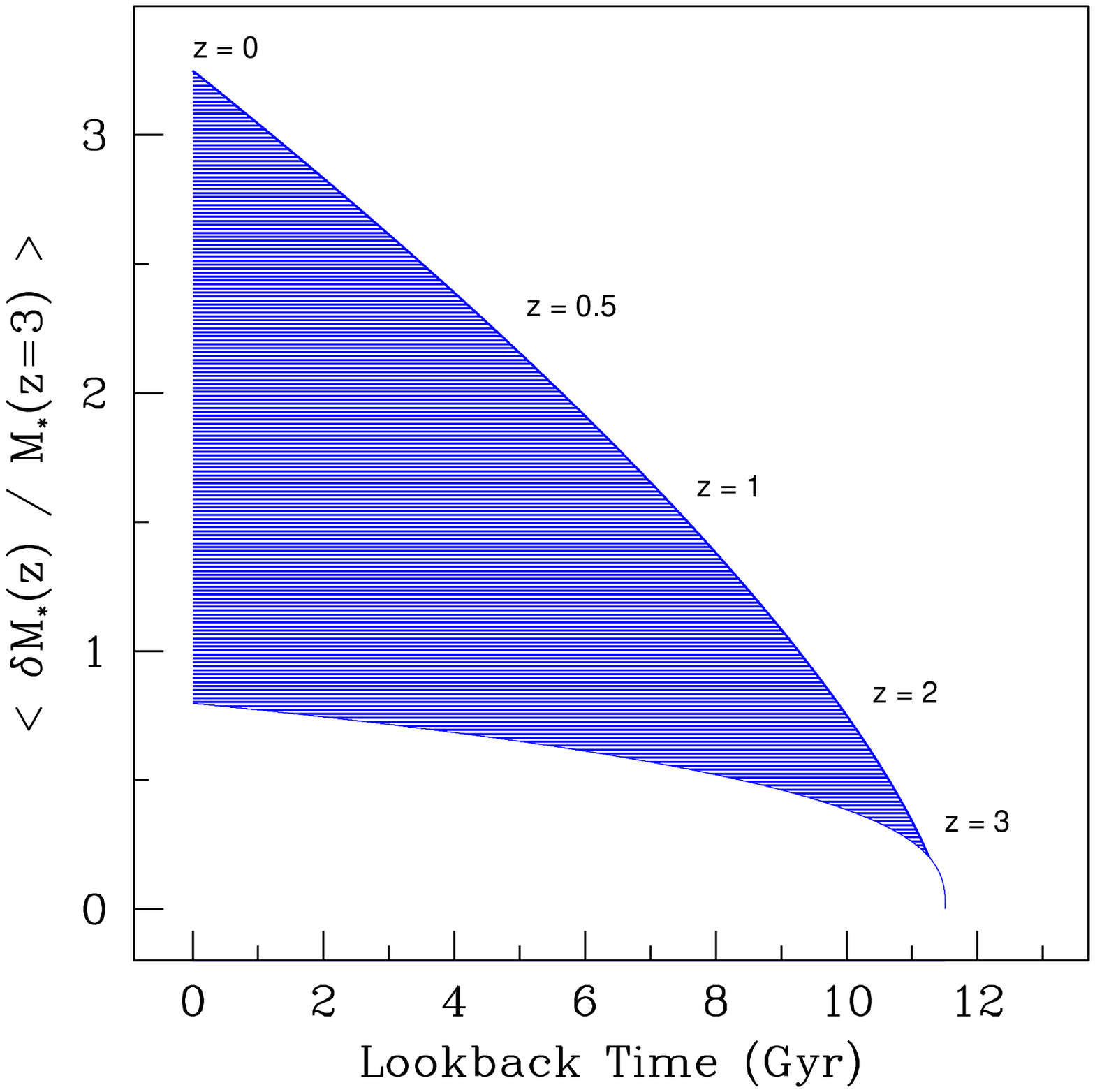}
\includegraphics[width=0.48\textwidth,height=0.48\textwidth]{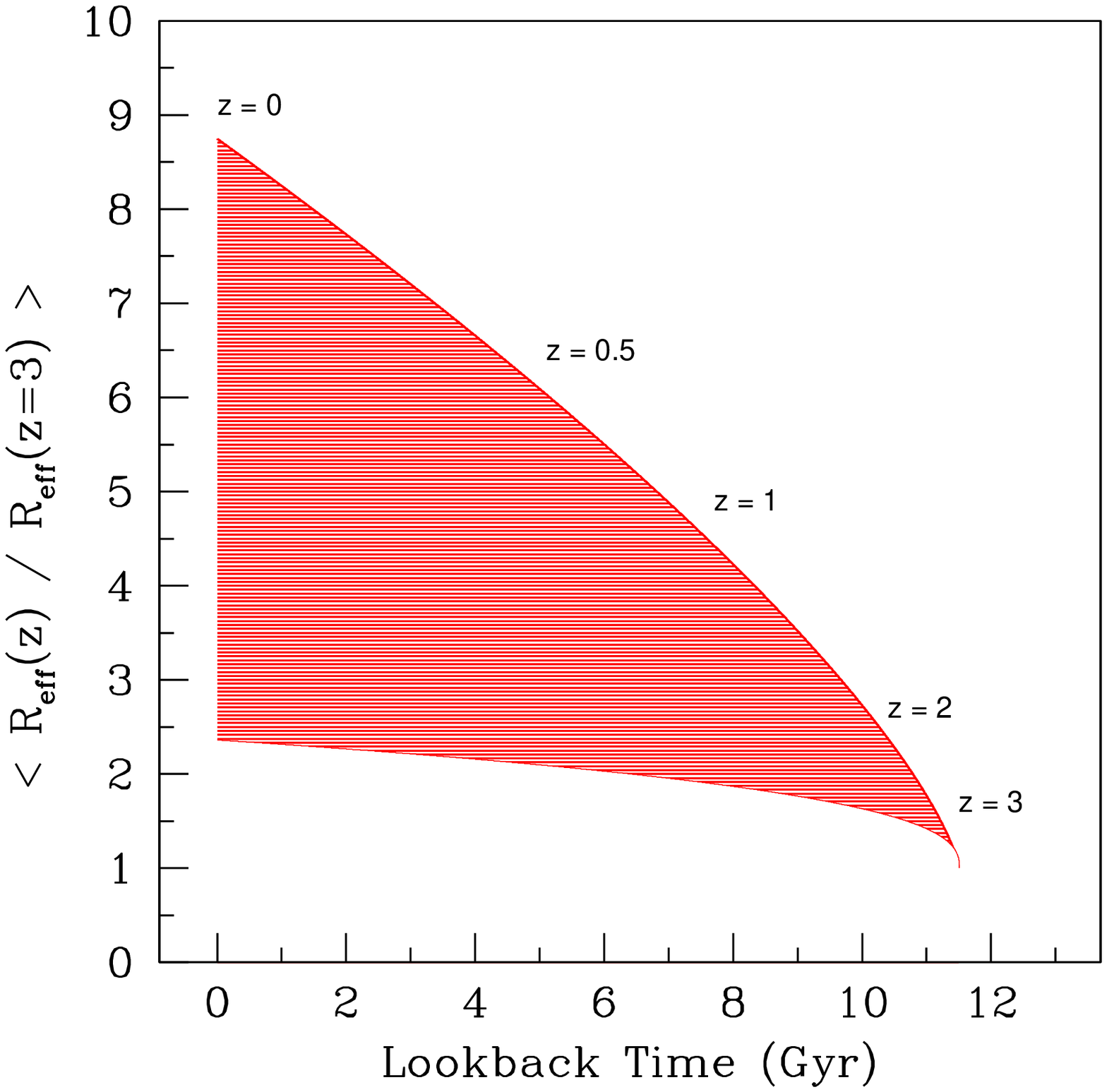}
\caption{Left panel: Permitted Stellar Mass Evolution. Right Panel: Permitted Size Evolution. The stellar mass and size evolution are plotted as a fuction of lookback time, overlayed are the redshifts at these times. Both plots are normalized by the initial z = 3 value, so the uncertainty increases towards the present. Only the lookback times 8 - 11.5 Gyr are probed directly by this study, with a wide range of probable functions used to extrapolate to the present (see \S 3.2.4), which is responsible for the flaring of these plots out to z = 0. The size evolution (right) is computed as the maximum allowable for the stellar mass growth (left). See \S 4.2 for full details.}
\end{center}
\end{figure*}

\section{Summary and Conclusions}

In this paper we study in detail the structures, morphologies, and total merger history evolution of a population of 80 extremely massive ($M_{*} > 10^{11} M_{\odot}$) galaxies at very high redshifts (1.7 $<$ z $<$ 3). We find that massive galaxies at high redshifts are frequently morphologically disturbed, with $\sim$ 1/4 meeting the criteria for a $CAS$ selected major merger (see \S 3.1.1). Furthermore, we note that close pair methods (with d $<$ 30 kpc) to deduce major mergers find close accord with morphological $CAS$ determined major mergers at all redshifts, up to z = 3, indicative of both approaches being sensitive to the underlying merger history, and, moreover, having similar mass ratio and timescale sensitivities.

We find that the `poorness' of GALFIT fitting (RFF) is positively correlated with the global asymmetry of the galaxy ($A$) for high redshift massive galaxies, see \S 3.1.3. Moreover, we find that galaxies with GALFIT fits defined as `poor' (RFF $>$ 0.05) are over four times more likely to be highly asymmetric systems (with $A$ $>$ 0.35) than those galaxies with `good' (RFF $<$ 0.05) fits, which suggests that Sersic index fitting can be used to select likely major merger candidates through analysis of the RFF.

We also investigate the minor merger properties of our sample of massive galaxies, down to a mass sensitivity threshold of $M_{*} = 10^{9} M_{\odot}$. In order to achieve this we utilise extremely deep and high resolution $H$ band imaging from the HST GOODS NICMOS Survey, which corresponds to rest frame optical wavelengths at the redshifts probed in this paper (z = 1.7 - 3). We deduce that there are in total $N_{m}$ = (4.5 +/- 2.9) / $\tau_{m}$ mergers from z = 3 to the present, which leads to massive galaxies growing in mass by a factor of $\sim$ 3 due to all mergers, with 1/2 of this mass increase being driven by minor mergers with galaxies of mass $M_{*} = 10^{9} - 10^{10.5} M_{\odot}$. We go on to suggest that the size increase over cosmic time of massive galaxies could be driven by these mergers, noting that the number of mergers we compute for massive galaxies (and their resultant approximate threefold mass increase) could correspond to a size increase of up to a factor of six via the Naab et al. (2009) model. This is examined in detail in \S 4.2.

In conclusion, we find that massive galaxies are more asymmetric at high redshifts, have many more minor and major mergers at early times than late times, and grow in mass threefold from z = 3 to the present due to merging alone. We also suggest that the size evolution of massive galaxies must be partly (or possibly entirely) driven via this galaxy merging.

\vspace{1cm}

\section*{Acknowledgements}

We thank Robert Chuter, Frazer Pearce and Nina Hatch for insightful discussions and advice on this work. We thank Emma Bradshaw for computational assistance with processing numerical solutions to complex integrals. We appreciate some very helpful suggestions and comments from the referee, Andrew Zirm. We also gratefully acknowledge funding from the STFC and the Leverhulme Trust to carry out this research.

\newpage

\section*{References}

\noindent Bauer A. et al., 2011, MNRAS in press, ArXiv:1106.2656
\\  Bertone S. \& Conselice C. J., 2009, MNRAS, 396, 2345
\\  Bezanson R. et al., 2009, ApJ, 697, 1290
\\  Binny J. \& Tremaine S., 2008, Princeton University Press,\\
\hspace*{0.5cm} ISBN 978-0-691-13026-2 
\\  Blakeslee J. P. et al., 2006, ApJ, 644, 30
\\  Bluck A. F. L., Conselice C. J., Bouwens R. J., Daddi E.,\\
\hspace*{0.5cm} Dickinson M.; Papovich C.; Yan H., 2009, MNRAS, \\
\hspace*{0.5cm} 394, 1956L
\\  Bluck A. F. L., Conselice C. J., Almaini O., Laird E., Nandra\\
\hspace*{0.5cm} K. \& Gruetzbauch R., 2011, MNRAS, 410, 1174
\\  Bridge C. R., Carlberg R. G. \& Sullivan M., 2010, ApJ, 709,\\
\hspace*{0.5cm} 1067
\\  Bruzual G. \& Charlot S., 2003, MNRAS, 344, 1000
\\  Buitrago F., Trujillo I., Conselice C. J., Bouwens R. J.,\\
\hspace*{0.5cm} Dickinson M., Yan H., 2008, ApJ, 687, 61
\\  Cappellari M. et al., 2009, 704, 34
\\  Carrasco, E. R., Conselice, C. J. \& Trujillo, I., 2010,\\
\hspace*{0.5cm} MNRAS, 405, 2253
\\  Cenarro A. J. \& Trujillo I., 2009, ApJ, 696, 43
\\  Ceverino D., Dekel A., Bournaud F., 2010, MNRAS,\\
\hspace*{0.5cm} 404, 2151
\\  Cimatti A. et al., 2008, A\&A, 482, 21
\\  Conselice C. J., 2003, ApJS, 147, 1
\\  Conselice C. J., 2006, ApJ, 639, 120
\\  Conselice C. J., 2009, MNRAS, 399, 16
\\  Conselice C. J., et al. 2003, AJ, 126, 1183
\\  Conselice C. J., et al. 2007, MNRAS, 381, 962
\\  Conselice C. J., et al. 2008, MNRAS, 386, 909
\\  Conselice C. J. et al., 2011a, MNRAS, 413, 80
\\  Conselice C. J. et al., 2011b, MNRAS, 1105, 2522
\\  Conselice C. J., Yang C., Bluck A. F. L., 2009, MNRAS,\\
\hspace*{0.5cm} 394, 1956
\\  Daddi E. et al., 2004, ApJ, 617, 746
\\  Daddi E. et al., 2007, ApJ, 670, 156
\\  De Propris et al., 2007, ApJ, 666, 212D
\\  De Vaucouleurs G., 1948, AnAp, 11, 247
\\  Dekel A. et al., 2009, Natur., 457, 451
\\  Elmegreen D. M. et al., 2011, ApJ, 737, 32
\\  Fan L., Lapi A., De Zotti G., Danese L., 2008, ApJ, 689, 101
\\  Forster-Schreiber N. M. et al., 2011, arXiv:1104.0248
\\  Franx M. et al., 2003, ApJ, 587, L93 
\\  Gr\"utzbauch R., Chuter R. W., Conselice C. J., Bauer A.,\\
\hspace*{0.5cm} Bluck A. F. L., Buitrago F., Mortlock A., 2011,\\
\hspace*{0.5cm} MNRAS, 412, 2361
\\  Gr\"utzbauch R. et al., 2011, MNRAS, in press,\\
\hspace*{0.5cm} arXiv:1108.0402
\\  Hopkins P. F., Bundy K., Murray N., Quataert E., Lauer\\
\hspace*{0.5cm} T. R., Ma C., 2009, MNRAS, 398, 898
\\  Hopkins P. F. et al., 2010, ApJ, 724, 915
\\  Hoyos C. et al., 2010, MNRAS, submitted
\\  Hubble E. P., 1926, ApJ, 64, 321
\\  Ibata R., Irwin M., Lewis G. F. \& Stolte A., 2001, ApJ,\\
\hspace*{0.5cm} 547, 133L
\\  Jogee S. et al., 2009, ApJ, 697, 1971
\\  Khochfar S. \& Burkert A., 2006, A\&A, 370, 902
\\  Khochfar S. \& Silk J., 2006, ApJ, 648, 21L
\\  Lopez-Sanjuan C. et al., 2010a, ApJ, 710, 1170
\\  Lopez-Sanjuan C., Balcells M., PŽrez-Gonz‡lez P. G.,\\
\hspace*{0.5cm} Barro G., Gallego J., Zamorano J., 2010b, A\&A,\\
\hspace*{0.5cm} 518, 20
\\  Lopez-Sanjuan C. et al., 2011, A\&A, 530, 20
\\  Lotz J. M. et al., 2008a, ApJ 672, 177
\\  Lotz J. M. et al., 2008b, MNRAS, in press\\
\hspace*{0.5cm} (arxiv:0805.1246v1)
\\  Lotz J. M., Jonsson P., Cox T. J. \& Primack J. R., 2010,\\
\hspace*{0.5cm} MNRAS, 404, 575L
\\  Lotz J. M. et al., 2011, ApJ in press, arXiv:1108.2508
\\  Man A. W. S., Toft S., Zirm A. W., Wuyts S., van der Wel A.,\\
\hspace*{0.5cm} 2011, accepted to ApJ, arXiv:1109.2895
\\  Mancini C. et al., 2010, MNRAS, 401, 933
\\  Metcalfe N., Shanks T., Weilbacher P. M., McCracken H. J.,\\
\hspace*{0.5cm} Fong R., Thompson D., 2006, MNRAS, 370, 1257
\\  Mortlock A., Conselice C. J., Bluck A. F. L., Bauer A. E.,\\
\hspace*{0.5cm} Gr\"utzbauch R., Buitrago F., Ownsworth J., 2011,\\
\hspace*{0.5cm} MNRAS, 413, 2845
\\  Naab T., Johansson P. H. \& Ostriker J. P., 2009, ApJ,\\
\hspace*{0.5cm} 699, 178L
\\  Onodera M. et al., 2010, ApJ, 715, 60
\\  Pannella M. et al., 2009, ApJ, 698, 116
\\  Papovich C. et al., 2006, ApJ, 640, 92
\\  Patton D. R. et al., 2000, ApJ, 536, 153
\\  Peng C. Y., Ho L. C., Impey C. D. \& Rix H., 2002, AJ, 124,\\
\hspace*{0.5cm} 266
\\  Rawat A. et al., 2008, ApJ, in press, (arXiv:0804.0078v1)
\\  Ricciardelli E., Trujillo I., Buitrago F., Conselice, C. J., 2010,\\
\hspace*{0.5cm} MNRAS, 406, 230
\\  Thompson R. I., Storrie-Lombardi L. J., Weymann R. J.,\\
\hspace*{0.5cm} Rieke M. J., Schneider G., Stobie E., Lytle D.,\\
\hspace*{0.5cm} 1999, AJ, 117, 17
\\  Trujillo I., Conselice C. J., Bundy K,, Cooper M. C.,\\
\hspace*{0.5cm} Eisenhardt P. \& Ellis R. S., 2007, MNRAS, 382,\\
\hspace*{0.5cm} 109
\\  Trujillo I., Cenarro A. J., de Lorenzo-Cáceres A., Vazdekis, A.,\\
\hspace*{0.5cm} de la Rosa I. G.\& Cava A., 2009, ApJ, 692, 118L
\\  Trujillo I., Ferreras I., de La Rosa I. G., 2011, MNRAS,\\
\hspace*{0.5cm} tmp. 938
\\  van Dokkum P. G., Kriek M., Franx M., 2009, Nature,\\
\hspace*{0.5cm} 460, 717
\\  Weinzirl T. et al., 2011, ApJ in press, arXiv:1107.2591
\\  Yan H. et al., 2004, ApJ, 616, 63

\newpage

\appendix
\section{Data Tables}

\begin{longtable}{@{}ccccccccc}
\caption{GNS Massive Galaxies: Structural and Close Pair Data}
\small
\label{tab1} \\
\hline
\hline
\multicolumn{1}{c}{GNS ID}&
\multicolumn{1}{c}{RA}&
\multicolumn{1}{c}{Dec}&
\multicolumn{1}{c}{z$_{phot}$}&
\multicolumn{1}{c}{Log($M_{*}$ [$M_{\odot}$])}&
\multicolumn{1}{c}{$N_{MP,corr}$}&
\multicolumn{1}{c}{$N_{TP,corr}$}&
\multicolumn{1}{c}{$C$}&
\multicolumn{1}{c}{$A$}\\
\hline
\endfirsthead

\multicolumn{9}{l}{{\tablename} \thetable{} --Continued}\\
\hline
\hline
\multicolumn{1}{c}{GNS ID}&
\multicolumn{1}{c}{RA}&
\multicolumn{1}{c}{Dec}&
\multicolumn{1}{c}{z$_{phot}$}&
\multicolumn{1}{c}{Log($M_{*}$ [$M_{\odot}$])}&
\multicolumn{1}{c}{$N_{MP,corr}$}&
\multicolumn{1}{c}{$N_{TP,corr}$}&
\multicolumn{1}{c}{$C$}&
\multicolumn{1}{c}{$A$}\\
\hline
\endhead

21 & 189.1354 & 62.1172 & 2.70 & 11.05 & 0.00 & 0.02 & 2.6 & 0.26  \\
43 & 189.1254 & 62.1155 & 2.20 & 11.08 & 0.00 & 0.00 & 3.1 & 0.35  \\
77 & 189.1326 & 62.1121 & 1.91 & 11.41 & 0.00 & 0.00 & 2.7 & 0.19  \\
227 & 189.1198 & 62.1355 & 2.07 & 11.20 & 0.00 & 0.03 & 3.1 & 0.27  \\
373 & 189.0587 & 62.1635 & 2.51 & 11.06 & 1.39 & 0.50 & 2.6 & 0.24  \\
552 & 189.0772 & 62.1510 & 1.93 & 11.33 & 0.00 & 1.42 & 2.9 & 0.20 \\
730 & 189.2512 & 62.1527 & 2.47 & 11.09 & 0.00 & 0.00 & 2.5 & 0.31  \\
840 & 189.1737 & 62.1673 & 1.93 & 11.30 & 0.00 & 0.97 & 2.7 & 0.17  \\
856 & 189.1788 & 62.1663 & 1.74 & 11.38 & 0.00 & 0.00 & 3.2 & 0.37  \\
999 & 189.1430 & 62.2335 & 1.98 & 11.17 & 0.00 & 0.00 & 2.2 & 0.27  \\
1129 & 189.5074 & 62.2717 & 2.37 & 11.19 & 0.00 & 0.00 & 2.9 & 0.40  \\
1144 & 189.5035 & 62.2700 & 2.07 & 11.13 & 0.00 & 0.00 & 2.9 & 0.25  \\
1257 & 189.2264 & 62.2922 & 2.02 & 11.00 & 0.00 & 0.46 & 2.4 & 0.30  \\
1394 & 189.3997 & 62.3453 & 2.04 & 11.36 & 0.00 & 0.00 & 2.8 & 0.20  \\
1533 & 189.3058 & 62.1791 & 2.56 & 11.13 & 1.75 & 3.02 & 3.0 & 0.27  \\
1666 & 189.2567 & 62.1962 & 2.36 & 11.38 & 0.73 & 0.22 & 2.5 & 0.24  \\
1769 & 189.2737 & 62.1871 & 1.95 & 11.32 & 1.59 & 0.00 & 3.5 & 0.45  \\
1826 & 189.0732 & 62.2613 & 2.20 & 11.28 & 0.51 & 1.17 & 2.3 & 0.32  \\
1942 & 189.2777 & 62.2546 & 2.51 & 11.03 & 1.65 & 0.88 & 2.6 & 0.32  \\
2049 & 189.3130 & 62.2047 & 2.40 & 11.21 & 1.25 & 1.30 & 2.3 & 0.34  \\
2066 & 189.3002 & 62.2034 & 2.80 & 11.35 & 1.21 & 2.21 & 2.6 & 0.30  \\
2083 & 189.3122 & 62.2016 & 2.72 & 11.28 & 0.44 & 0.43 & 2.3 & 0.38  \\
2282 & 189.3069 & 62.2626 & 2.30 & 11.13 & 0.00 & 0.00 & 2.5 & 0.40  \\
2411 & 189.0479 & 62.1761 & 2.10 & 11.17 & 0.00 & 0.00 & 2.3 & 0.15  \\
2564 & 189.2110 & 62.2488 & 1.83 & 11.13 & 0.00 & 0.00 & 2.7 & 0.43  \\
2678 & 189.0475 & 62.1486 & 2.50 & 11.53 & 0.58 & 0.00 & 0.0 & 0.42  \\
2734 & 189.0423 & 62.1463 & 2.60 & 11.03 & 0.00 & 0.07 & 1.2 & 0.14  \\
2764 & 189.0525 & 62.1433 & 2.20 & 11.15 & 0.00 & 0.00 & 2.5 & 0.26  \\
2837 & 189.0944 & 62.2750 & 2.30 & 11.39 & 0.37 & 1.61 & 3.0 & 0.27  \\
2902 & 189.0913 & 62.2677 & 2.00 & 11.22 & 0.00 & 0.00 & 2.8 & 0.34  \\
2965 & 189.0799 & 62.2449 & 2.80 & 11.13 & 0.25 & 1.29 & 2.7 & 0.29  \\
3036 & 189.0869 & 62.2377 & 2.10 & 11.11 & 0.57 & 1.84 & 1.9 & 0.48  \\
3126 & 189.1304 & 62.1661 & 2.10 & 11.09 & 0.00 & 0.42 & 2.9 & 0.36  \\
3250 & 189.2291 & 62.1385 & 2.30 & 11.12 & 0.00 & 0.00 & 2.5 & 0.29  \\
3387 & 189.2942 & 62.3472 & 1.83 & 11.01 & 0.47 & 0.00 & 2.3 & 0.41  \\
3422 & 189.2808 & 62.3442 & 2.80 & 11.09 & 0.00 & 0.00 & 2.4 & 0.27  \\
3582 & 189.0987 & 62.1693 & 2.40 & 11.18 & 0.00 & 0.00 & 3.4 & 0.35  \\
3629 & 189.1829 & 62.2725 & 2.10 & 11.28 & 1.69 & 2.11 & 2.6 & 0.25  \\
3766 & 189.2056 & 62.3226 & 2.10 & 11.19 & 0.00 & 0.00 & 2.5 & 0.29  \\
3818 & 189.2022 & 62.3171 & 1.75 & 11.41 & 0.77 & 0.00 & 3.4 & 0.27  \\
3822 & 189.2198 & 62.3169 & 2.20 & 11.06 & 0.00 & 0.00 & 2.9 & 0.16  \\
3970 & 189.3319 & 62.2059 & 2.34 & 11.16 & 0.40 & 0.45 & 2.9 & 0.33  \\
4033 & 189.4641 & 62.2440 & 2.07 & 11.04 & 0.74 & 1.20 & 2.7 & 0.30  \\
4121 & 189.4564 & 62.2332 & 1.92 & 11.20 & 0.00 & 0.07 & 2.4 & 0.60  \\
4239 & 188.9812 & 62.1738 & 2.20 & 11.17 & 0.00 & 0.00 & 2.7 & 0.23  \\
4282 & 53.0938 & -27.8013 & 2.60 & 11.08 & 0.00 & 1.07 & 2.9 & 0.35  \\
4301 & 53.0923 & -27.8031 & 2.40 & 11.33 & 0.00 & 0.00 & 2.3 & 0.21  \\
4353 & 53.1011 & -27.8086 & 1.97 & 11.21 & 0.00 & 0.00 & 2.7 & 0.22  \\
4400 & 53.1013 & -27.7117 & 2.30 & 11.10 & 0.73 & 0.16 & 2.8 & 0.23  \\
4434 & 53.0976 & -27.7153 & 2.14 & 11.06 & 0.54 & 0.00 & 2.7 & 0.30  \\
4557 & 53.0892 & -27.7601 & 2.27 & 11.21 & 0.65 & 0.02 & 2.8 & 0.27  \\
4706 & 53.1231 & -27.8033 & 2.35 & 11.10 & 0.00 & 0.00 & 2.8 & 0.31  \\
4754 & 53.1200 & -27.8082 & 2.00 & 11.50 & 0.00 & 0.00 & 2.4 & 0.20  \\
4882 & 53.1718 & -27.8257 & 1.74 & 11.10 & 0.00 & 0.00 & 2.7 & 0.33 \\
4941 & 53.2300 & -27.8508 & 1.83 & 11.01 & 0.00 & 0.00 & 0.0 & 0.00  \\
5171 & 53.0632 & -27.6997 & 2.40 & 11.01 & 0.00 & 0.00 & 2.3 & 0.36  \\
5282 & 53.0860 & -27.7096 & 2.10 & 11.09 & 0.00 & 1.02 & 2.8 & 0.24  \\
5372 & 53.1255 & -27.8864 & 2.90 & 11.08 & 0.00 & 0.00 & 2.2 & 0.37  \\
5445 & 53.1245 & -27.8932 & 2.50 & 11.45 & 0.10 & 0.17 & 2.5 & 0.35  \\
5524 & 53.1333 & -27.9029 & 2.58 & 11.17 & 0.00 & 0.00 & 2.4 & 0.40  \\
5533 & 53.1289 & -27.9037 & 2.79 & 11.10 & 0.00 & 1.06 & 2.4 & 0.27  \\
5764 & 53.2252 & -27.8738 & 2.65 & 11.27 & 0.00 & 0.00 & 2.9 & 0.40  \\
5853 & 53.0507 & -27.7138 & 2.41 & 11.44 & 2.10 & 0.31 & 3.1 & 0.12  \\
5933 & 53.0542 & -27.7217 & 2.30 & 11.26 & 0.30 & 0.00 & 2.9 & 0.43  \\
6035 & 53.0555 & -27.8739 & 1.90 & 11.32 & 0.00 & 0.11 & 3.2 & 0.31  \\
6114 & 53.0656 & -27.8788 & 2.24 & 11.11 & 0.69 & 1.04 & 2.8 & 0.22  \\
6220 & 53.0716 & -27.8436 & 1.90 & 11.01 & 0.55 & 0.00 & 2.7 & 0.16  \\
6352 & 53.0773 & -27.8596 & 1.96 & 11.08 & 0.00 & 0.00 & 2.7 & 0.14  \\
6468 & 53.1385 & -27.6718 & 2.80 & 11.45 & 0.36 & 0.00 & 2.6 & 0.16  \\
6575 & 53.0355 & -27.6901 & 2.50 & 11.41 & 0.00 & 0.00 & 2.5 & 0.29  \\
6584 & 53.0260 & -27.6909 & 2.20 & 11.52 & 0.00 & 0.00 & 2.9 & 0.27  \\
6584 & 53.0261 & -27.6910 & 1.99 & 11.07 & 0.00 & 0.00 & 2.9 & 0.25  \\
6876 & 53.0400 & -27.6852 & 2.50 & 11.45 & 0.34 & 0.00 & 2.5 & 0.29  \\
7090 & 53.0578 & -27.8335 & 2.70 & 11.68 & 0.00 & 0.00 & 3.1 & 0.24  \\
7156 & 53.1178 & -27.9109 & 2.69 & 11.17 & 0.00 & 0.00 & 3.9 & 0.13  \\
7321 & 53.1161 & -27.8719 & 2.07 & 11.00 & 0.00 & 0.00 & 2.7 & 0.31  \\
7425 & 53.1272 & -27.8345 & 1.81 & 11.22 & 0.00 & 0.00 & 2.9 & 0.27  \\
7677 & 53.1830 & -27.7090 & 1.76 & 11.57 & 0.00 & 0.00 & 3.4 & 0.34  \\
7970 & 53.0282 & -27.7788 & 2.30 & 11.15 & 0.00 & 1.61 & 2.2 & 0.24  \\
8140 & 53.1410 & -27.7667 & 1.91 & 11.22 & 0.00 & 0.00 & 2.9 & 0.26  \\
8214 & 53.1623 & -27.7121 & 2.14 & 11.16 & 0.64 & 0.01 & 2.9 & 0.31  \\
\hline
\end{longtable}

\noindent In the above table we present data on the fundamental properties, close pair statistics, and structures of the 80 massive ($M_{*} > 10^{11} M_{\odot}$) galaxies found within the GNS. $N_{MP,corr}$ is the statistically corrected major pair fraction of each galaxy, with $N_{TP,corr}$ the statistically corrected total pair fraction of each galaxy with paired galaxies with $M_{*} > 10^{9} M_{\odot}$. $A$ is the asymmetry of the massive galaxy, with $C$ being its concentration. Typical errors are: $< A_{err} >$ = $^{+}_{-}$0.05; $< C_{err} >$ = $^{+}_{-}$0.26; and $\sim$ 0.3 dex for stellar mass.

\noindent Full data catalogs are available at http://www.nottingham.ac.uk/$\sim$ppzgns/index.html, where all computed values and initial error estimations are publically available. Please see Conselice et al. (2011a) for a companion to these data products.

\newpage

\begin{table}
 \caption{Close Pair Method Test: Model Spectroscopy form the Millennium Simulation}
 \label{tab2}
 \begin{tabular}{@{}ccccccc}
  \hline
\hline
\hspace{2cm}   Run ID   &   N$_{hosts}$   &   N$_{total}$   &   $f_{m}$ (stats)   &   $f_{m}$ ($<$ 500 km/s)   &   $f_{m}$ ($<$ 1000 km/s)   &   $f_{m}$ ($<$ 1500 km/s) \hspace{2cm}  \\
\hline
\hspace{2.5cm}  1 & 44 & 30738 & 0.149+/-0.022 & 0.068+/-0.010 & 0.114+/-0.017 & 0.159+/-0.024 \hspace{2.5cm} \\
\hspace{2.5cm}  2 & 43 & 31295 & 0.247+/-0.038 & 0.210+/-0.032 & 0.233+/-0.036 & 0.280+/-0.043 \hspace{2.5cm}  \\
\hspace{2.5cm}  3 & 82 & 31921 & 0.110+/-0.012 & 0.061+/-0.007 & 0.098+/-0.010 & 0.146+/-0.016 \hspace{2.5cm}  \\
\hspace{2.5cm}  4 & 49 & 32501 & 0.161+/-0.023 & 0.121+/-0.017 & 0.167+/-0.024 & 0.198+/-0.028 \hspace{2.5cm} \\
\hspace{2.5cm}  5 & 62 & 33770 & 0.152+/-0.019 & 0.097+/-0.012 & 0.113+/-0.014 & 0.168+/-0.021 \hspace{2.5cm} \\
\hspace{2.5cm}  6 & 84 & 31948 & 0.189+/-0.020 & 0.123+/-0.013 & 0.166+/-0.018 & 0.221+/-0.024 \hspace{2.5cm} \\
\hspace{2.5cm}  7 & 54 & 34376 & 0.161+/-0.022 & 0.019+/-0.003 & 0.093+/-0.013 & 0.167+/-0.023 \hspace{2.5cm} \\
\hspace{2.5cm}  8 & 65 & 31727 & 0.191+/-0.024 & 0.070+/-0.009 & 0.142+/-0.018 & 0.161+/-0.020 \hspace{2.5cm}  \\
\hspace{2.5cm}  mean & 60 & 32285 & 0.170+/-0.040 & 0.095+/-0.050 & 0.142+/-0.045 & 0.188+/-0.044 \hspace{2.5cm}  \\
\hline
\end{tabular}

\vspace{1cm}

\noindent This table presents the results from a series of eight Millennium Simulation run tests on the efficacy of the Close Pair method presented in this paper and Bluck et al. (2009). Presented here are close pair fractions deduced statistically, and through a spectroscopic-like method for comaprison. The final row presents the mean properties across the varying tests. The models used were taken from the Kitzbichler et al. (2006) view of the De Lucia et al. (2006) semianalytic galaxy catalog for the Millennium simulation. Please see http://www.g-vo.org/Millennium for full details and to access the mock catalogs used here for further investigation. The errors on indiviually calculated merger fractions are simple Poisson counting errors, with the errors presented for the mean values being the standard deviation about the mean across the eight runs performed.

\end{table}

\end{document}